\documentclass[12pt,preprint,number,sort&compress,lontitle,times]{elsarticle}
\usepackage{natbib}
\usepackage{graphicx}
\usepackage{epsfig}
\usepackage{lineno}
\usepackage{multirow}
\usepackage[latin1]{inputenc}
\usepackage[T1]{fontenc}
\usepackage{textcomp}

\begin{document}
\begin{frontmatter}
\title{Search for First Harmonic Modulation\\ 
in the Right Ascension Distribution of Cosmic Rays\\ 
Detected at the Pierre Auger Observatory}
\author{
{\bf The Pierre Auger Collaboration} \\
P.~Abreu$^{72}$, 
M.~Aglietta$^{55}$, 
E.J.~Ahn$^{88}$, 
I.F.M.~Albuquerque$^{17,\: 88}$, 
D.~Allard$^{31}$, 
I.~Allekotte$^{1}$, 
J.~Allen$^{91}$, 
P.~Allison$^{93}$, 
J.~Alvarez Castillo$^{65}$, 
J.~Alvarez-Mu\~{n}iz$^{79}$
% ...
\\
$^{1}$ Centro At\'{o}mico Bariloche and Instituto Balseiro (CNEA-
UNCuyo-CONICET), San Carlos de Bariloche, Argentina \\
% ...
$^{17}$ Universidade de S\~{a}o Paulo, Instituto de F\'{\i}sica, S\~{a}o 
Paulo, SP, Brazil \\
% ...
$^{31}$ Laboratoire AstroParticule et Cosmologie (APC), 
Universit\'{e} Paris 7, CNRS-IN2P3, Paris, France \\
% ...
$^{55}$ Istituto di Fisica dello Spazio Interplanetario (INAF),
 Universit\`{a} di Torino and Sezione INFN, Torino, Italy \\
% ...
$^{65}$ Universidad Nacional Autonoma de Mexico, Mexico, D.F., 
Mexico \\
% ...
$^{72}$ LIP and Instituto Superior T\'{e}cnico, Lisboa, Portugal \\
% ...
$^{79}$ Universidad de Santiago de Compostela, Spain \\
% ...
$^{88}$ Fermilab, Batavia, IL, USA \\
% ...
$^{91}$ New York University, New York, NY, USA \\
% ...
$^{93}$ Ohio State University, Columbus, OH, USA \\
% ...
($\star$) Corresponding author~: auger\_pc@fnal.gov \\
($\dagger$) Deceased \\
(a) at Konan University, Kobe, Japan\\
}

\begin{abstract}
We present the results of searches for dipolar-type anisotropies in different energy 
ranges above $2.5\times 10^{17}$~eV with the surface detector array of the Pierre Auger 
Observatory, reporting on both the phase and the amplitude measurements of the 
first harmonic modulation in the right-ascension distribution. 
Upper limits on the amplitudes are obtained, which provide 
the most stringent bounds at present, being below 2\%  at 99\% $C.L.$ for EeV energies. 
We also compare our results to those of previous experiments as well as with 
some theoretical expectations.
\end{abstract}
\end{frontmatter}


\section{Introduction}

The large-scale distribution of the arrival directions of Ultra-High Energy Cosmic Rays 
(UHECRs) is, together with the spectrum and the mass composition, an important observable in 
attempts to understand their nature and origin. The \textit{ankle}, a hardening of the 
energy spectrum of UHECRs  located at $E\simeq4\,$EeV~\cite{linsley-ankle,lawrence,nagano,
bird,auger-spectrum}, where 1~${\rm EeV}\equiv 10^{18}$~eV, is presumed to be either 
the signature of the transition from galactic to extragalactic UHECRs~\cite{linsley-ankle}, 
or the distortion of a proton-dominated extragalactic spectrum due to $e^{\pm}$ pair 
production of protons with the photons of the Cosmic Microwave Background (CMB)~\cite{hillas,berezinsky}. 
If cosmic rays with energies below the ankle have a galactic origin, their escape from 
the Galaxy might generate a dipolar large-scale pattern as seen from the Earth. The 
amplitude of such a pattern is difficult to predict, as it depends on the assumed 
galactic magnetic field and the charges of the particles as well as the distribution 
of sources. Some estimates, in which the galactic cosmic rays are mostly heavy, show that 
anisotropies at the level of a few percent are nevertheless expected in the EeV 
range~\cite{ptuskin,roulet1}. Even for isotropic extragalactic cosmic rays, a dipole 
anisotropy may exist due to our motion with respect to the frame of extragalactic 
isotropy.  This \emph{Compton-Getting effect}~\cite{compton} has been measured with 
cosmic rays of much lower energy at the solar frequency~\cite{eastop, groom} as 
a result of our motion relative to the frame in which they have no bulk motion.

Since January 2004, the surface detector (SD) array of the Pierre Auger Observatory has 
collected a large amount of data. The statistics accumulated in the $1\,$EeV energy range 
allows one to be sensitive to intrinsic anisotropies with amplitudes down to the 1\%  level. 
This requires determination of the exposure of the sky at a corresponding accuracy (see 
Section~\ref{sec:exposure}) as well as control of the systematic uncertainty of the 
variations in the counting rate of events induced by the changes of the atmospheric conditions 
(see Section~\ref{sec:weather}). After carefully correcting these experimental effects, we 
present in Section~\ref{sec:results} searches for first harmonic modulations in right-ascension 
based on the classical Rayleigh analysis~\cite{linsley-fh} slightly modified to account for 
the small variations of the exposure with right ascension. 

Below $E\simeq1\,$EeV, the detection efficiency of the array depends on zenith angle and 
composition, which amplifies detector-dependent variations in the counting rate.  
Consequently, our results below 1~EeV are derived using simple event counting rate 
differences between Eastward and Westward directions~\cite{ew}.  That technique using 
relative rates allows a search for anisotropy in right ascension without requiring any 
evaluation of the detection efficiency.

From the results presented in this work, we derive in Section~\ref{sec:discussion}
upper limits on modulations in right-ascension of UHECRs and discuss some of their 
implications. 

\section{The Pierre Auger Observatory and the data set}
\label{sec:auger}

The southern site of the Pierre Auger Observatory~\cite{auger-nim} is located in Malarg\"{u}e, 
Argentina, at latitude 35.2$^\circ\,$S, longitude 69.5$^\circ\,$W and mean altitude 1400 
meters above sea level. Two complementary techniques are used to detect extensive air
showers initiated by UHECRs~: a \textit{surface detector array} and a \textit{fluorescence
detector}. The SD array consists of 1660 water-Cherenkov detectors covering an area of
about 3000~km$^2$ on a triangular grid with 1.5~km spacing, allowing electrons, photons
and muons in air showers to be sampled at ground level with a duty cycle of almost 100\%. 
In addition, the atmosphere above the SD array is observed during clear, dark nights by
24 optical telescopes grouped in 4 buildings. These detectors observe the longitudinal 
profile of air showers by detecting the fluorescence light emitted by nitrogen molecules 
excited by the cascade. 

The data set analysed here consists of events recorded by the surface detector from 1 January 
2004 to 31 December 2009. During this time, the size of the Observatory increased from 154 to 
1660 surface detector stations. We consider in the present analysis events\footnote{A 
comprehensive description of the identification of shower candidates detected at the SD array 
of the Pierre Auger Observatory is given in reference~\cite{auger-acc}.} with reconstructed 
zenith angles smaller than 60$^\circ$ and satisfying a fiducial cut requiring that the six 
neighbouring detectors in the hexagon surrounding the detector with the highest 
signal were active when the event was recorded. Throughout this article, based on this fiducial 
cut, any active detector with six active neighbours will be defined as 
an \emph{unitary cell}~\cite{auger-acc}. It ensures both a good quality 
of event reconstruction and a robust estimation of the exposure of the SD array, which is then 
obtained in a purely geometrical way. The analysis reported here is restricted 
to selected periods to eliminate unavoidable problems associated to the construction 
phase, typically in the data acquisition and the communication system or due to hardware 
instabilities~\cite{auger-acc}. These cuts restrict the duty cycle to $\simeq 85$\%. Above the 
energy at which the detection efficiency saturates, 3~EeV~\cite{auger-acc}, the exposure 
of the SD array is 16,323~km$^2$~sr~yr for six years used in this analysis.

The event direction is determined from a fit to the arrival times of the shower front at the 
SD. The precision achieved in this reconstruction depends upon the accuracy on the GPS 
clock resolution and on the fluctuations in the time of arrival of the first particle~\cite{ang-res}. 
The angular resolution is defined as the angular aperture around the arrival directions of 
cosmic rays within which 68\% of the showers are reconstructed. At the lowest observed energies, 
events trigger as few as three surface detectors. The angular resolution of events having such a 
low multiplicity is contained within $2.2^\circ$, which is quite sufficient to perform searches 
for large-scale patterns in arrival directions, and reaches $\sim 1^\circ$ for events with 
multiplicities larger than five~\cite{ang-res2}.

The energy of each event is determined in a two-step procedure. First, using the constant
intensity cut method, the shower size at a reference distance of $1000\,$m, $S(1000)$, is 
converted to the value $S_{38^\circ}$ that would have been expected had the shower arrived 
at a zenith angle 38$^\circ$. Then, $S_{38^\circ}$ is converted to energy 
using a calibration curve based on the fluorescence telescope measurements~\cite{auger-prl}. 
The uncertainty in $S_{38^\circ}$ resulting from the adjustment of the shower size, the 
conversion to a reference angle, the fluctuations from shower-to-shower and the calibration 
curve amounts to about 15\%. The absolute energy scale is given by the fluorescence 
measurements and has a systematic uncertainty of 22\%~\cite{auger-prl}.

\section{The exposure of the surface detector}
\label{sec:exposure}

\begin{figure}[!t]
  \centering					 
  \includegraphics[width=7.5cm]{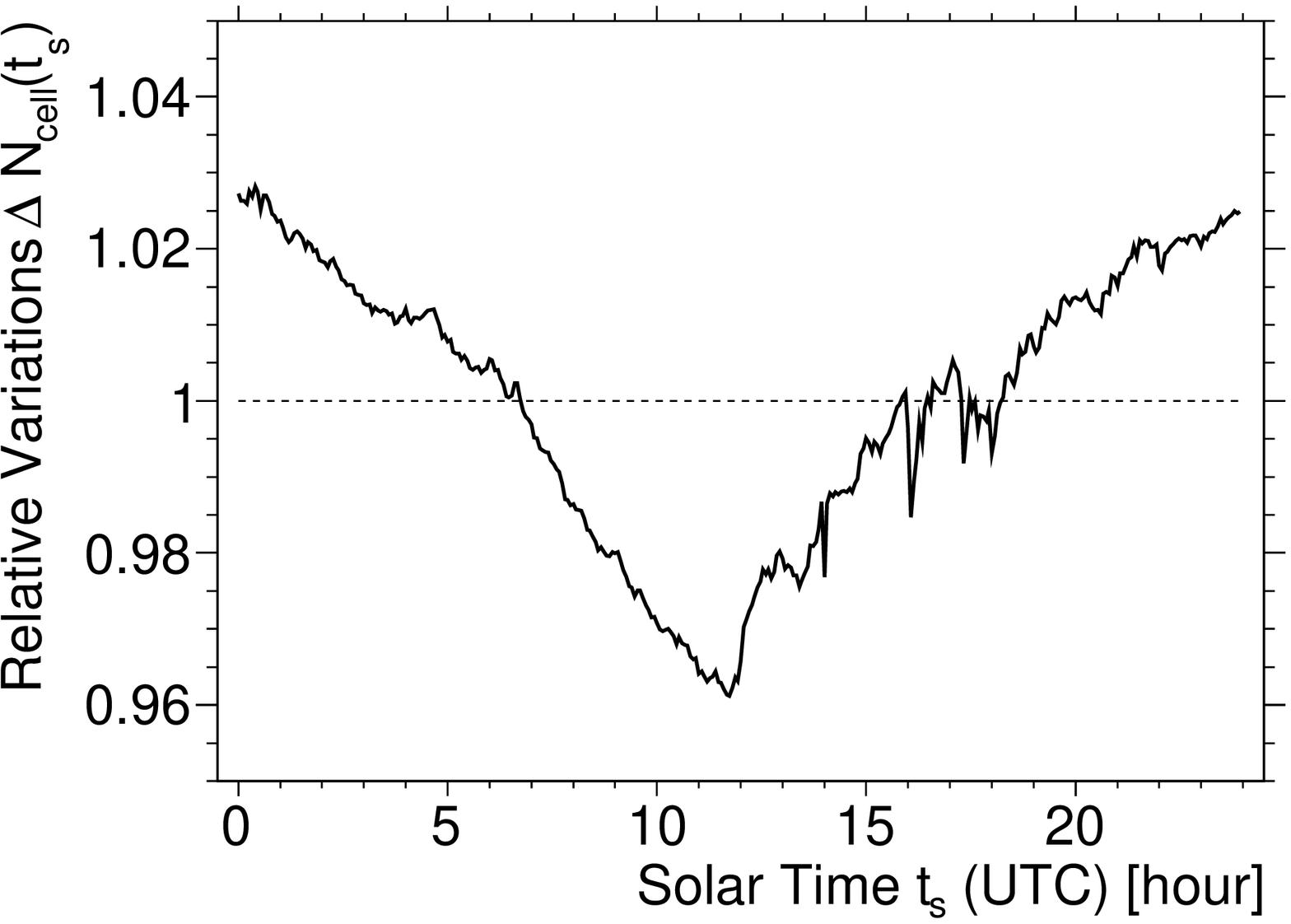}
  \includegraphics[width=7.5cm]{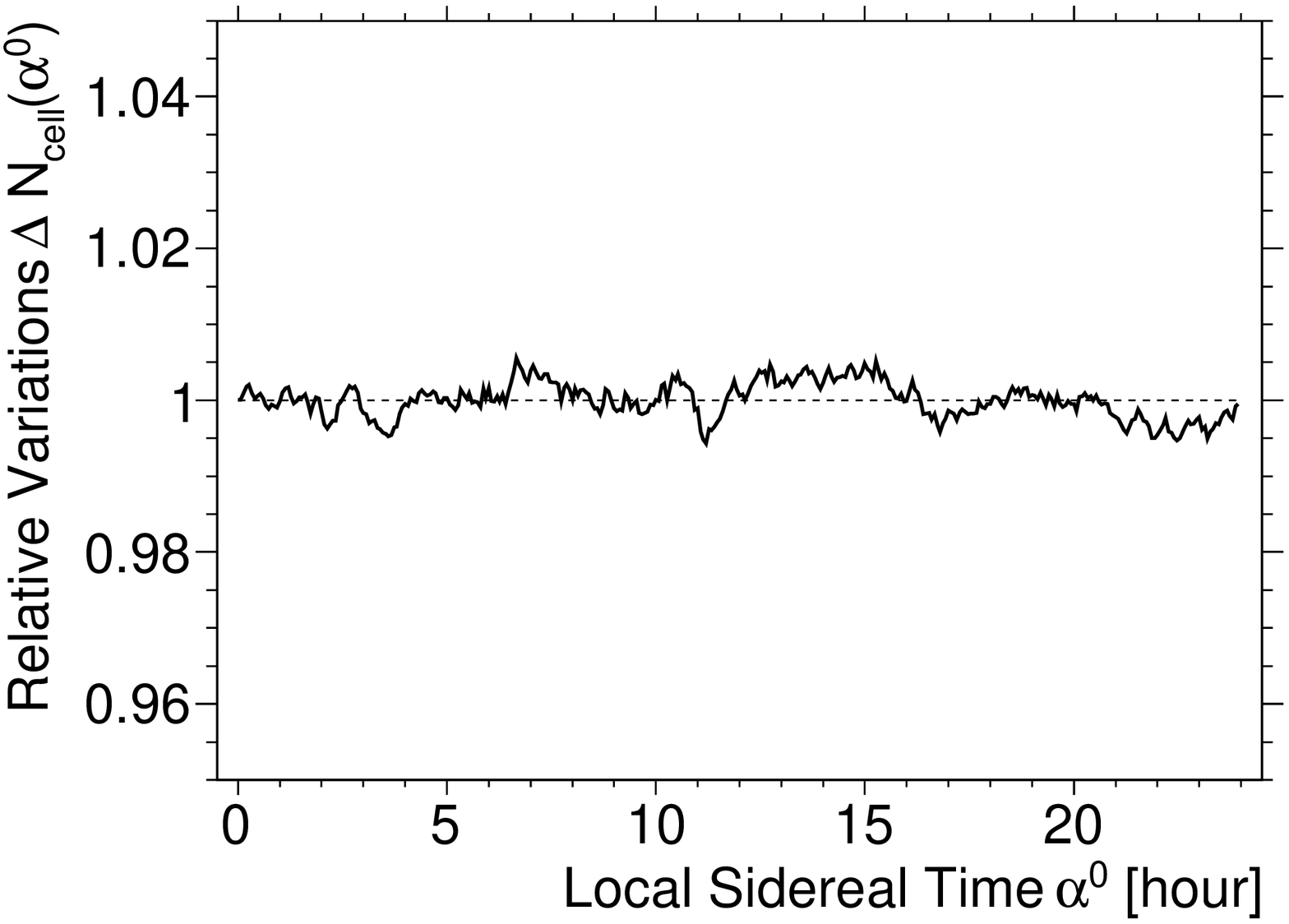}
  \caption{\small{Relative variation of the integrated number of unitary cells as a 
function of the solar hour of the day in UTC (left panel), and as a function of the local sidereal 
time (right panel).}}
  \label{N_h}
\end{figure}

The instantaneous exposure $\omega(t,\theta,\phi,S_{38^\circ})$ of the SD array at the 
time $t$ as a function of the incident zenith and azimuth\footnote{The angle $\phi$ is 
the azimuth relative to the East direction, measured counterclockwise.} angles ($\theta,\phi$) 
and shower size $S_{38^\circ}$ is given by~:
\begin{equation}
\label{eqn:omega}
\omega(t,\theta,\phi,S_{38^\circ})=n_{\mathrm{cell}}(t) \times a_{\mathrm{cell}} \cos{\theta}\times\epsilon(S_{38^\circ},\theta,\phi),
\end{equation}
where $a_{\mathrm{cell}} \cos{\theta}$ is the projected surface of a unitary cell under 
the incidence zenith angle $\theta$, $n_{\mathrm{cell}}(t)$ is the number of unitary cells  
at time $t$, and $\epsilon(S_{38^\circ},\theta,\phi)$ is the directional detection 
efficiency at size parameter $S_{38^\circ}$ under incidence angles ($\theta,\phi$). 
The conversion from $S_{38^\circ}$ to the energy $E$, which accounts for the changes of atmospheric 
conditions, will be presented in the next section. 

The number of unitary cells $n_{\mathrm{cell}}(t)$ is recorded every second using the trigger 
system of the Observatory and reflects the array growth as well as the dead periods of each 
detector. It ranges from $\simeq60$ (at the begining of the data taking in 2004) to 
$\simeq1200$ (from the middle of 2008). From Eqn.~\ref{eqn:omega}, it is apparent that 
$n_{\mathrm{cell}}(t)$ is the only time-dependent quantity entering in the definition of the 
instantaneous exposure, modulating within any integrated solid angle the expected number of 
events as a function of time. For any periodicity $T$, the total number of unitary cells  
$N_{\mathrm{cell}}(t)$ as a function of time $t$ within a period and summed over all periods, 
and its associated relative variations $\Delta N_{\mathrm{cell}}(t)$ are obtained from~:
\begin{equation}
N_{\mathrm{cell}}(t)=\sum_j n_{\mathrm{cell}}(t+jT), \hspace{1cm} 
\Delta N_{\mathrm{cell}}(t)=\frac{N_{\mathrm{cell}}(t)}{\left<N_{\mathrm{cell}}(t)\right>},
\end{equation}
with $\left<N_{\mathrm{cell}}(t)\right>=1/T\,\int_0^T dt\,N_{\mathrm{cell}}(t)$.

A genuine dipolar anisotropy in the right ascension distribution of the events induces 
a modulation in the distribution of the time of arrival of events with a period equal to 
one sidereal day. A sidereal day indeed corresponds to the time it takes for the Earth to 
complete one rotation relative to the vernal equinox. It is approximately $T_{sid} =$ 23 hours, 
56 minutes, 4.091 seconds. Throughout this article, we denote by $\alpha^0$ the local sidereal 
time and express it in hours or in radians, as appropriate. For practical reasons, $\alpha^0$ 
is chosen so that it is always equal to the right ascension of the zenith at the center of the 
array.

On the other hand, a dipolar modulation \textit{of experimental origin} in the distribution of the
time of arrival of events with a period equal to one solar day may induce a spurious dipolar 
anisotropy in the right ascension distribution of the events. Hence, it is essential to 
control $\Delta N_{\mathrm{cell}}(t)$ to account for the variation of the 
exposure in different directions. We show $\Delta N_{\mathrm{cell}}(t)$ in Fig.~\ref{N_h} 
in 360 bins of 4 minutes at these two time scales of particular interest~: 
the solar one $T=T_{sol}$ = 24 hs (left panel), and the sidereal one $T=T_{sid}$ (right 
panel). A clear diurnal variation is apparent 
on the solar time scale showing an almost dipolar modulation with an amplitude
of $\simeq 2.5$\%. This is due to both the working times of the construction phase of the 
detector and to the outage of some batteries of the surface detector stations during nights. 
When averaged over 6 full years, this modulation is almost totally smoothed out on the sidereal 
time scale as seen in the right panel of Fig.~\ref{N_h}. This distribution will be used in 
section \ref{sec:RayAna} to weight the events when estimating the Rayleigh parameters. 

From the instantaneous exposure, it is straightforward to compute the integrated exposure 
either in local coordinates $\omega(\theta,\phi,\alpha^0)$ by replacing 
$n_{\mathrm{cell}}(t)$ by $\Delta N_{\mathrm{cell}}(\alpha^0)$ in Eqn.~\ref{eqn:omega}, 
or in celestial coordinates $\omega(\alpha,\delta)$ by expressing the zenith angle $\theta$ 
in terms of the equatorial right ascension $\alpha$ and declination $\delta$ through~:
\begin{equation}
\label{eqn:theta}
\cos{\theta}=\sin{\ell_{\mathrm{site}}}\sin{\delta}+\cos{\ell_{\mathrm{site}}}\cos{\delta}\cos{(\alpha-\alpha^0)},
\end{equation}
(where $\ell_{\mathrm{site}}$ is the Earth latitude of the site) and then by 
integrating Eqn.~\ref{eqn:omega} over time. Besides, let us also mention
that to account for the spatial extension of the surface detector array making the latitude
of the site $\ell_{\mathrm{site}}$ varying by $\simeq 0.4^\circ$, the celestial 
coordinates $(\alpha,\delta)$ of the events are calculated by transporting the showers
to the "center" of the Observatory site. 

\section{Influence of the weather effects}
\label{sec:weather}

Changes in the atmospheric pressure $P$ and air density $\rho$ have been shown to affect the 
development of extensive air showers detected by the surface detector array and these changes 
are reflected in the temporal variations of shower size at a fixed energy~\cite{auger-weather}.
To eliminate these variations, the procedure used to convert the observed signal into energy 
needs to account for these atmospheric effects. This is performed by relating the signal at 1~km 
from the core, $S(1000)$, measured at the actual density $\rho$ and pressure $P$, to the one 
$\tilde{S}(1000)$ that would have been measured at reference values $\rho_0$ and $P_0$, 
chosen as the average values at Malarg\"ue, \textit{i.e.} 
$\rho_0=1.06$ kg~m$^{-3}$ and $P_0=862$~hPa~\cite{auger-weather}~:
\begin{equation}
\tilde{S}(1000)=\left[1-\alpha_P(\theta)(P-P_0)-\alpha_\rho(\theta)(\rho_d-\rho_0)-\beta_\rho(\theta)(\rho-\rho_d)\right]S(1000),
\label{sweather}
\end{equation}
where $\rho_d$ is the average daily density at the time the event was recorded. The measured 
coefficients $\alpha_\rho=(-0.80\pm0.02)\,$kg$^{-1}$m$^3$, $\beta_\rho=(-0.21\pm0.02)\,$kg$^{-1}$m$^3$ 
and $\alpha_P=(-1.1\pm0.1)\,10^{-3}\,$hPa$^{-1}$ reflect respectively the impact of the variation 
of air density (and thus temperature) at long and short time scales, and of the variation of 
pressure on the shower sizes~\cite{auger-weather}. It is worth pointing out that air density 
coefficients are here predominant relative to the pressure one. The zenithal dependences of these 
parameters, that we use in the following, were also studied in reference~\cite{auger-weather}.
It is this reference signal $\tilde{S}(1000)$ which has to be
converted, using the constant intensity cut method, to the signal size $\tilde{S}_{38^\circ}$ 
and finally to energy. For convenience, we denote hereafter the uncorrected (corrected) 
shower size $S_{38^\circ}$ ($\tilde{S}_{38^\circ}$) simply by $S$ ($\tilde{S}$).

Carrying out such energy corrections is important for the study of large scale anisotro-
pies. 
Above 3~EeV, the rate of events $R$ per unit time above a given \emph{uncorrected} size threshold 
$S_{th}$, and in a given zenith angle bin, is modulated by changes of atmospheric conditions~:
\begin{equation}
R(>S_{th})\propto\int_{S_{th}}^{\infty}dS\,\frac{dJ}{d\tilde{S}}\,\frac{d\tilde{S}}{dS}\propto\bigg[1+(\gamma_S-1)\,\alpha_\xi\,\Delta\xi\bigg]\,\int_{S_{th}}^{\infty}dS\,S^{-\gamma_S},
\label{def_rate_sat}
\end{equation}
where hereafter $\xi$ generically denotes $P$, $\rho$ or $\rho_d$, and where we have
adopted for the differential flux $dJ/dS$ a power law with spectral index $\gamma_S$.
Hence, under changes of atmospheric parameters $\Delta\xi$, the following relative 
change in the rate of events is expected~:
\begin{eqnarray}
\frac{1}{R}\frac{dR(>S_{th})}{d\xi}&\simeq&(\gamma_S-1)\,\alpha_\xi.
\label{rel_rate_sat}
\end{eqnarray}
Over a whole year, this spurious modulation is partially compensated in sidereal time, 
though not in solar time. In addition, a seasonal variation of the modulation of the daily 
counting rate induces sidebands at both the sidereal and anti-sidereal\footnote{The anti-sidereal 
time is a fictitious time scale symmetrical to the sidereal one with respect to the solar one 
and that reflects seasonal influences~\cite{far-sto}. It corresponds to a fictitious year of
$\simeq 364$ days.} frequencies. This may lead to 
misleading measures of anisotropy if the amplitude of the sidebands significantly stands 
out above the background noise~\cite{far-sto}. Correcting energies for weather effects, the
net correction in the first harmonic amplitude in \emph{sidereal time} turns out to be only 
of $\simeq 0.2\%$ for energy thresholds greater than 3~EeV, thanks to large cancellations 
taking place when considering the large time period used in this study.

\begin{figure}[!t]
  \centering					 
  \includegraphics[width=10cm]{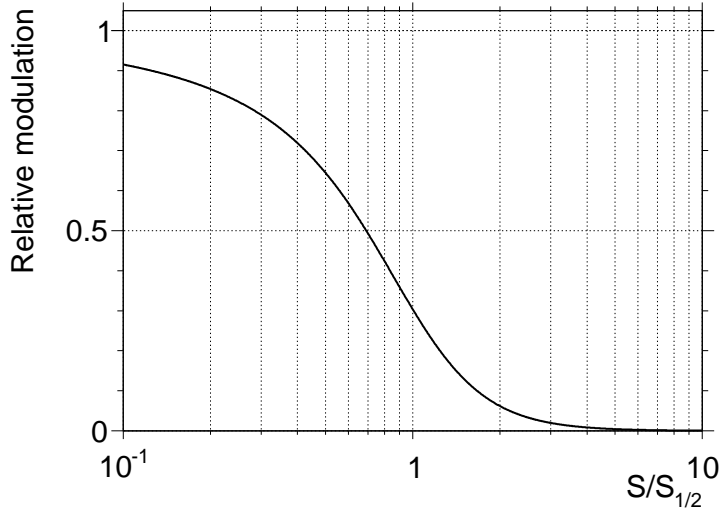}
  \caption{\small{Relative modulation of the rates above a given corrected signal size due 
to the variations of the detection efficiency under changes of atmospheric conditions, 
relatively to the factor modulating the rate of events above the corresponding uncorrected 
signal size in units of $S_{1/2}$.}}
  \label{rel_mod}
\end{figure}

In addition to the energy determination, weather effects can also affect the detection 
efficiencies for showers with energies below 3~EeV, for the detection of which the surface 
array is not fully efficient. Changes of the shower signal size due to changes of weather conditions
$\Delta\xi$ imply that showers are detected with the efficiency associated to the 
\emph{observed} signal size $S$, which is related at first order to the one 
associated to the \emph{corrected} signal size through~:
\begin{equation}
\epsilon(S)\simeq\epsilon(\tilde{S})+(S-\tilde{S})\frac{d\epsilon(S)}{dS}\bigg|_{S=\tilde{S}}\simeq\epsilon(\tilde{S})+\alpha_\xi\Delta\xi\,\tilde{S}\,\frac{d\epsilon(S)}{dS}\bigg|_{S=\tilde{S}},
\label{epsilon_variation}
\end{equation}
where we have made use of Eqn.~\ref{sweather}. 
The second term modulates the observed rate of events, even after the correction of the
signal sizes. Indeed, the rate of events $R$ above a given \emph{corrected} signal size 
threshold $\tilde{S}_{th}$ is now the integration of the cosmic ray spectrum weighted 
by the corresponding detection efficiency expressed in terms of the \emph{observed} signal 
size $S$~:
\begin{equation}
R(>\tilde{S}_{th})\propto\int_{\tilde{S}_{th}}^{\infty}d\tilde{S}\,\bigg[\epsilon(\tilde{S})+\alpha_\xi\Delta\xi\,\tilde{S}\,\frac{d\epsilon(S)}{dS}\bigg|_{S=\tilde{S}}\bigg]\,\frac{dJ}{d\tilde{S}}.
\label{def_rate}
\end{equation}
Hence, the relative change in the rate of events under changes in the atmosphere becomes~:
\begin{eqnarray}
\frac{1}{R}\frac{dR(>\tilde{S}_{th})}{d\xi}\simeq\frac{\alpha_\xi}{R}\int_{\tilde{S}_{th}}^{\infty}d\tilde{S}\,\tilde{S}\,\frac{dJ}{d\tilde{S}}\,\frac{d\epsilon(S)}{dS}\bigg|_{S=\tilde{S}},
\label{def_rel_rate}
\end{eqnarray}
which, after an integration by parts and at first order in $\alpha_\xi$, leads to~:
\begin{eqnarray}
\frac{1}{R}\frac{dR(>\tilde{S}_{th})}{d\xi}&\simeq&(\gamma_S-1)\,\alpha_\xi\,\bigg[1-\frac{\epsilon(\tilde{S}_{th})\int_{\tilde{S}_{th}}^{\infty}d\tilde{S}\,\tilde{S}^{-\gamma_S}}{\int_{\tilde{S}_{th}}^{\infty}d\tilde{S}\,\epsilon(\tilde{S})\,\tilde{S}^{-\gamma_S}}\bigg].
\label{rel_rate}
\end{eqnarray}
The expression in brackets gives the additional modulations (in units of the weather
effect modulation $(\gamma_s-1)\alpha_\xi$ when the detection efficiency is saturated) 
due to the variation of the detection efficiency. Note that this expression is less than 1 for any 
rising function $\epsilon$ satisfying $0\leq\epsilon(S)\leq 1$, and reduces to 0, as expected, 
when $\epsilon(\tilde{S}_{th})=1$. As a typical example, we show in 
Fig.~\ref{rel_mod} the expected modulation amplitude as a function of $S$ by adopting a 
reasonable detection efficiency function of the form $\epsilon(S)=S^3/\big[S^3+S_{1/2}^3\big]$ 
where the value of $S_{1/2}$ is such that $\epsilon(S_{1/2})=0.5$. This relative amplitude is 
about 0.3 for $S=S_{1/2}$, showing that for this signal size threshold the remaining modulation 
of the rate of events after the signal size corrections is about $0.3\times(\gamma_s-1)\alpha_\xi$. 
The value of $S_{1/2}$ being such that it corresponds to $\simeq 0.7$~EeV in terms of energy, 
it turns out that within the current statistics the Rayleigh analysis of arrival directions can
be performed down to threshold energies of 1$\,$EeV by \emph{only} correcting the energy 
assignments.

\section{Analysis methods and results}
\label{sec:results}

\subsection{Overview of the analyses}

The distribution in right ascension of the flux of CRs arriving at a detector can be 
characterised by the amplitudes and phases of its Fourier expansion, 
$I(\alpha) = I_0 (1 + r \cos(\alpha -\varphi) + r' \cos(2(\alpha - \varphi')) + \dots)$. 
Our aim is to determine the first harmonic amplitude $r$ and its phase $\varphi$.
To account for the non-uniform exposure of the SD array, we perform two different 
analyses.

\subsubsection{Rayleigh analysis weighted by exposure}
\label{sec:RayAna}
 
Above 1$\,$EeV, we search for the first harmonic modulation in right ascension 
by applying the classical Rayleigh formalism~\cite{linsley-fh} 
slightly modified to account for the non-uniform exposure to different parts of the sky.
This is achieved by weighting each event with a factor inversely proportional to the 
integrated number of unitary cells at the local sidereal time of the event (given by 
the right panel histogram of Fig.~\ref{N_h})~\cite{sommers,roulet2}:
\begin{equation}
\label{eqn:fh}
a=\frac{2}{\mathcal{N}}\sum_{i=1}^N w_i\cos{\alpha_i}, 
\hspace{1cm}b=\frac{2}{\mathcal{N}}\sum_{i=1}^N w_i\sin{\alpha_i},
\end{equation}
where the sum runs over the number of events $N$ in the considered energy range, the weights
are given by $w_i\equiv [\Delta N_{\mathrm{cell}}(\alpha^0_i)]^{-1}$ and the normalisation 
factor is $\mathcal{N}=\sum_{i=1}^Nw_i$. The estimated amplitude $r$ and phase $\varphi$ are 
then given by~:
\begin{equation}
\label{eqn:fh2}
r=\sqrt{a^2+b^2}, \hspace{1cm}\varphi=\arctan\frac{b}{a}.
\end{equation}
As the deviations from an uniform right ascension exposure are small,
the probability $P(>r)$ that an amplitude equal or larger than $r$ arises from an isotropic 
distribution can be approximated by the cumulative distribution function of the Rayleigh 
distribution $P(>r)=\exp{(-k_0)}$, where $k_0=\mathcal{N}r^2/4$. 

\subsubsection{East-West method}

Below 1$\,$EeV, due to the variations of the event counting rate arising from Eqn.~\ref{rel_rate}, 
we adopt the differential \textit{East-West method}~\cite{ew}. Since the instantaneous exposure 
of the detector for Eastward and Westward events is the same\footnote{The global tilt of the 
array of $\simeq 0.2^\circ$, that makes it slightly asymmetric, is here negligible - see 
sub-section~\ref{addcrosschecks}.}, with both sectors being equally affected by the instabilities 
of the detector and the weather conditions, the difference between the event counting rate 
measured from the East sector, $I_E(\alpha^0)$, and the West sector, $I_W(\alpha^0)$, 
allows us to remove at first order the direction independent effects of experimental 
origin without applying any correction, though at the cost of a reduced sensitivity. 
Meanwhile, this counting difference is directly related to the right ascension modulation 
$r$ by (see Appendix)~:
\begin{equation}
I_E(\alpha^0)-I_W(\alpha^0) = - \frac{N}{2\pi}\frac{2\left<\sin{\theta}\right>}
{\pi \langle \cos \delta \rangle} r \sin (\alpha^0 -\varphi).
\end{equation}
The amplitude $r$ and phase $\varphi$ can thus be calculated from the arrival times of each set 
of $N$ events using the standard first harmonic analysis~\cite{linsley-fh} slightly modified to 
account for the subtraction of the Western sector to the Eastern one. The Fourier 
coefficients $a_{EW}$ and $b_{EW}$ are thus defined by~:
\begin{eqnarray}
\label{eqn:fhew}
a_{EW} = \frac{2}{N}\sum_{i=1}^{N} \cos{(\alpha^0_i+\zeta_i)},\hspace{1cm}
b_{EW} = \frac{2}{N}\sum_{i=1}^{N} \sin{(\alpha^0_i+\zeta_i)},
\end{eqnarray}
where $\zeta_i$ equals 0 if the event is coming from the East or $\pi$ if coming from the West
(so as to effectively subtract the events from the West direction).
This allows us to recover the amplitude $r$ and the phase $\varphi_{EW}$ from
\begin{equation}
\label{eqn:fh2ew}
r=\sqrt{a_{EW}^2+b_{EW}^2}, \hspace{1cm}\varphi_{EW}=
\arctan{\left(\frac{b_{EW}}{a_{EW}}\right)}.
\end{equation}
Note however that $\varphi_{EW}$, being the phase corresponding to the maximum 
in the differential of the East and West fluxes, is related to $\varphi$ in 
Eqn.~\ref{eqn:fh2} through $\varphi = \varphi_{EW}+\pi/2$. As in the previous analysis, 
the probability $P(>r)$ that an amplitude equal or larger than $r$ arises from an isotropic 
distribution is obtained by the cumulative distribution function of the Rayleigh distribution 
$P(>r)=\exp{(-k_0^{EW})}$, where 
$k_0^{EW}=(2\left<\sin{\theta}\right>/\pi\langle\cos\delta\rangle)^2\times Nr^2/4$. For the 
values of $\left<\sin{\theta}\right>$ and $\langle\cos\delta\rangle$ of the events used in 
this analysis, the first factor in the expression for $k_0^{EW}$ is 0.22. Then, comparing 
it with the expression for $k_0$ in the standard Rayleigh analysis, it is seen that 
approximately four times more events are needed in the East-West method to attain the same 
sensitivity to a given amplitude $r$.

\subsection{Analysis of solar, anti-sidereal, and random frequencies}
\label{sec:solar-antisid}

\begin{figure}[!t]
  \centering					 
  \includegraphics[width=10cm]{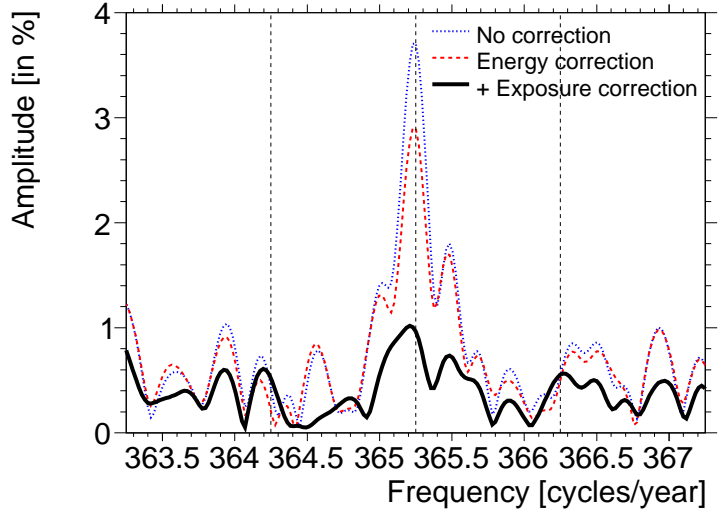}
  \caption{\small{Amplitude of the Fourier modes as a function of the frequency above 1$\,$EeV. 
Thin dotted curve~: before correction of energies and exposure. Dashed curve~: after correction 
of energies but before correction of exposure. Thick curve~: After correction of energies and 
exposure. Dashed vertical lines from left to right~: anti-sidereal, solar and sidereal frequencies.}}
  \label{spectralanalysis}
\end{figure}

The amplitude $r$ corresponds to the value of the Fourier transform of the arrival time distribution 
of the events at the sidereal frequency. This can be generalised to other frequencies by performing 
the Fourier transform of the modified time distribution~\cite{bil-let}~:
\begin{equation}
\label{eqn:modtimes}
\tilde{\alpha}^0_i=\frac{2\pi}{T_{sid}}t_i+\alpha_i-\alpha^0_i.
\end{equation}

Such a generalisation is helpful for examining an eventual residual spurious modulation
after applying the Rayleigh analysis after the corrections discussed in Sections 
\ref{sec:exposure} and \ref{sec:weather}.
The amplitude of the Fourier modes when considering all events above 1$\,$EeV are shown 
in Fig.~\ref{spectralanalysis} as a function of the frequency in a window centered on the 
solar one (indicated by the dashed line at 365.25~cycles/year). The thin dotted curve is obtained 
without accounting for the variations of the exposure and without accounting for the weather 
effects. The large period of time analysed here, over 6 years, allows us to resolve the 
frequencies at the level of $\simeq 1/6$~cycles/year. This induces a large decoupling of the 
frequencies separated by more than this resolution~\cite{bil-let}. In particular, as the 
resolution is less than the difference between the solar and the (anti-)sidereal frequencies 
(which is of 1~cycle/year), this explains why the large spurious modulations standing out from 
the background noise around the solar frequency are largely averaged out at both the sidereal 
and anti-sidereal frequencies even without applying any correction.
The impact of the correction of the energies discussed in Section~\ref{sec:weather} is evidenced 
by the dashed curve, which shows a reduction of $\simeq30$\% of the spurious modulations 
within the resolved solar peak. In addition, when accounting also for the exposure variation at 
each frequency, the solar peak is reduced at a level close to the statistical noise, as evidenced 
by the thick curve. Results at the solar and the anti-sidereal frequencies are collected 
in Tab.~\ref{tab}. 

\begin{table*}[h]
\begin{center}
\begin{tabular}{c|c|c|c|c}
 & $r_{\mathrm{solar}}$[\%] & $P(>r_{\mathrm{solar}})$[\%] & $r_{\mathrm{anti-sid}}$[\%] & $P(>r_{\mathrm{anti-sid}})$[\%] \\
\hline
\hline
no correction & 3.7 & $\simeq 2\,10^{-37}$ & 0.36 & 43 \\
energy corrections & 2.9 & $\simeq 4\,10^{-23}$ & 0.15 & 85  \\
+exposure correction & 0.96 & 0.2 & 0.49 & 19 
\end{tabular}
\caption{\small{Amplitude and corresponding probability to get a larger amplitude from an 
isotropic distribution at both the solar and the anti-sidereal frequencies for events with 
energies $>1\,$EeV.}}
\label{tab}
\end{center}
\end{table*}%

\begin{figure}[!t]
  \centering					 
  \includegraphics[width=7.5cm]{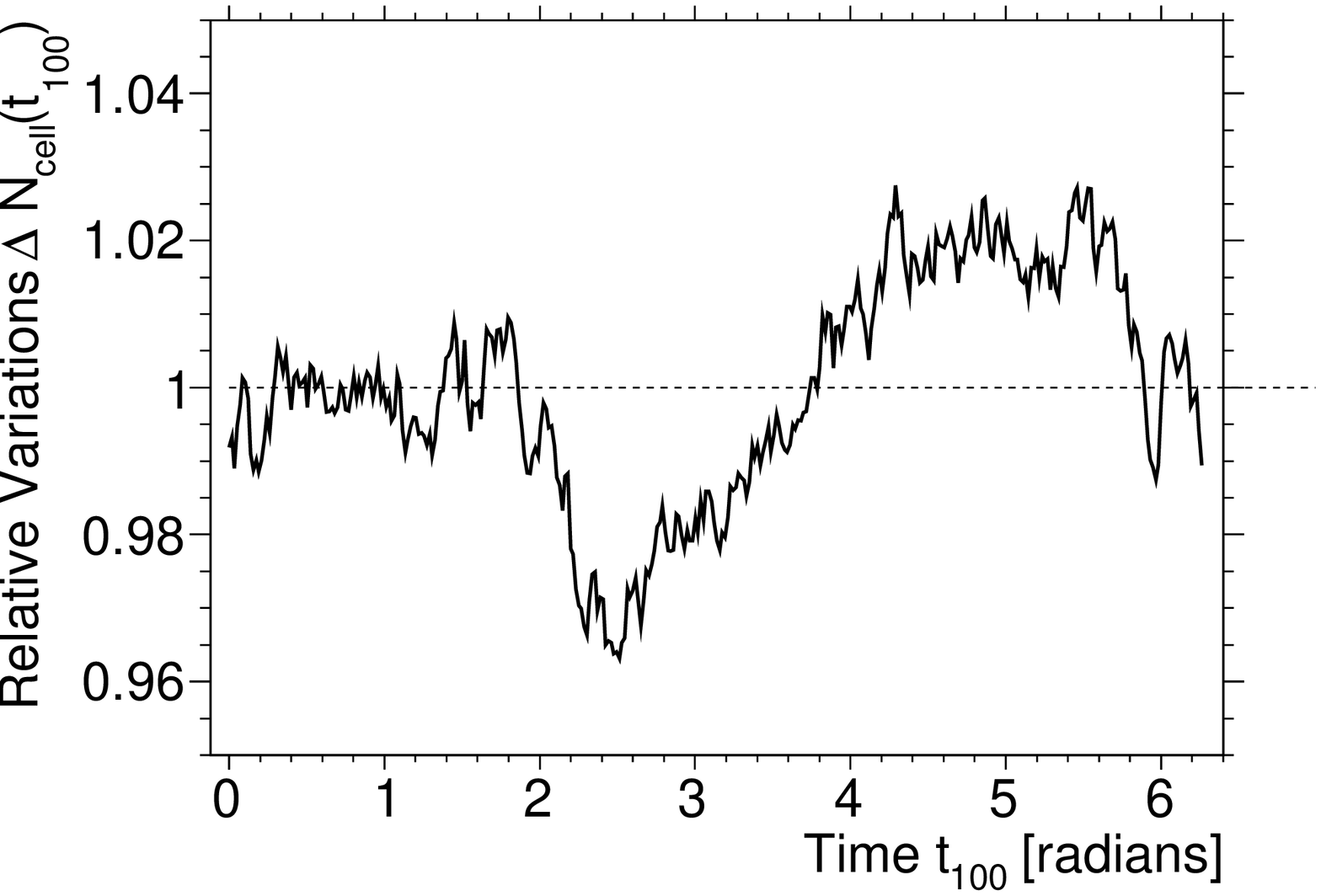}
  \includegraphics[width=7.5cm]{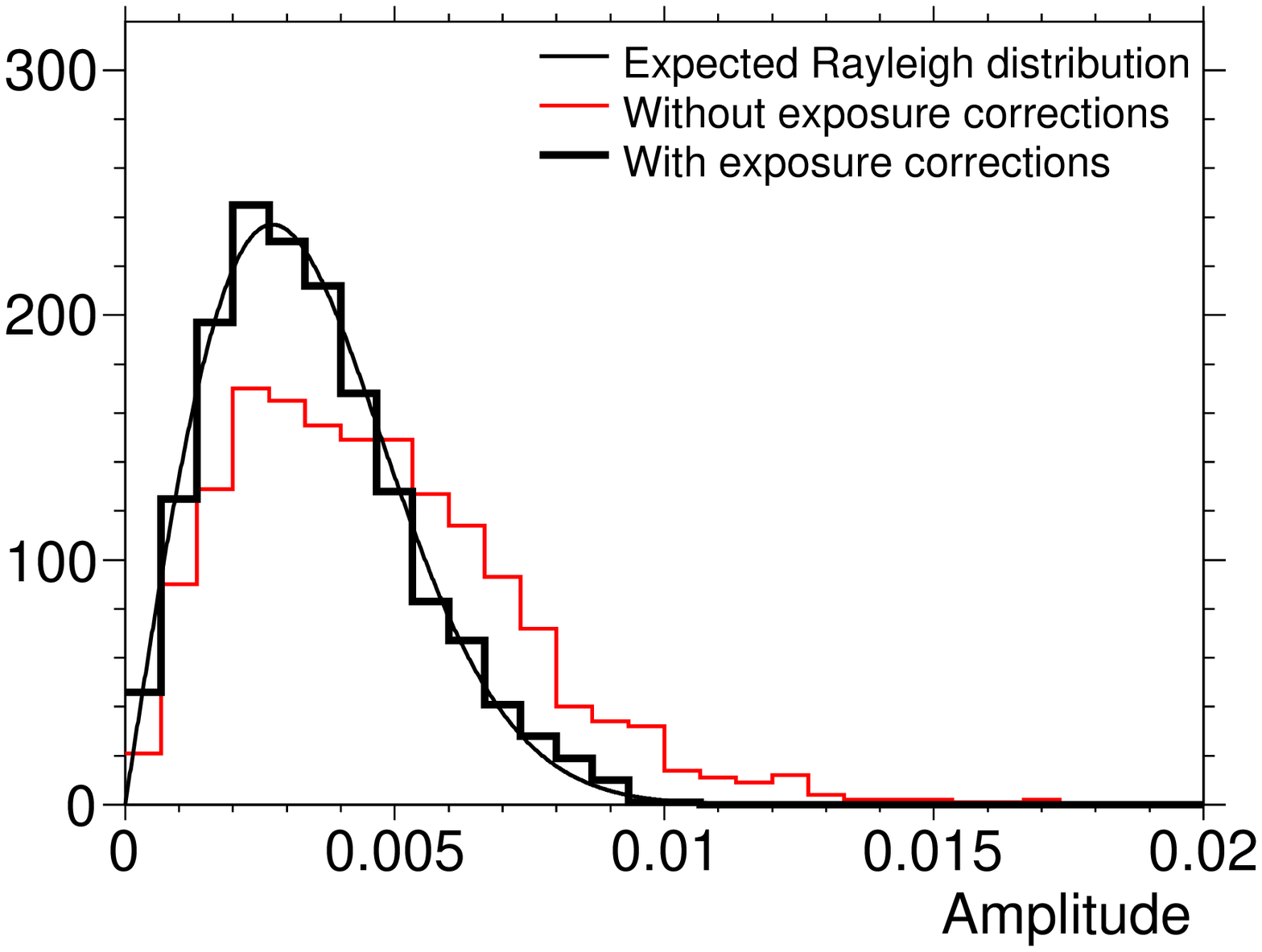}
  \caption{\small{Left~: Relative variation of the integrated number of unitary cells as a 
function of the time $t_{100}$, where the time scale is such that corresponding frequency is 
100~cycles/year. Right~: Rayleigh analysis above 1~EeV for 1600 random frequencies ranging 
from 100 to 500~cycles/year. Thin histogram~: analysis without accounting for the exposure 
variations. Thick histogram~: analysis accounting for the exposure variations. Smooth curve~: 
expected Rayleigh distribution.}}
  \label{fourierspectrum}
\end{figure}

To provide further evidence of the relevance of the corrections introduced to account 
for the non-uniform exposure, it is worth analysing on a statistical basis the behaviour 
of the reconstructed amplitudes at different frequencies (besides the anti-sidereal/solar/sidereal 
ones). In particular, as the number of unitary cells $n_{\mathrm{cell}}$ has increased 
from $\simeq$ 60 to $\simeq$ 1200 over the 6 years of data taking, an automatic
increase of the variations of $\Delta N_{\mathrm{cell}}(t)$ is expected at large time 
periods. This expectation is illustrated in the left panel of Fig.~\ref{fourierspectrum}, 
which is similar to Fig.~\ref{N_h} but at a time periodicity $T\simeq87.5\,$h, corresponding 
to a low frequency of 100~cycles/year. The size of the modulation is of the order 
of the one observed in Fig.~\ref{N_h} at the solar frequency. In the right panel of 
Fig.~\ref{fourierspectrum}, the results of the Rayleigh analysis applied above 1~EeV 
to 1600 random frequencies ranging from 100 to 500~cycles/year are shown by histograming 
the reconstructed amplitudes. The thin one is obtained \emph{without} accounting for the 
variations of the exposure~: it clearly deviates from the expected Rayleigh distribution 
displayed in the same graph. Once the exposure variations are accounted for through the 
weighting procedure, the thick histogram is obtained, now in agreement with the expected 
distribution. Note that in both cases the energies are corrected for weather effects, but 
the impact of these effects is marginal when considering such random frequencies. This 
provides additional support that the variations of the counting rate induced by the variations 
of the exposure are under control through the monitoring of $\Delta N_{\mathrm{cell}}(t)$. 

\begin{figure}[!t]
  \centering					 
  \includegraphics[width=10cm]{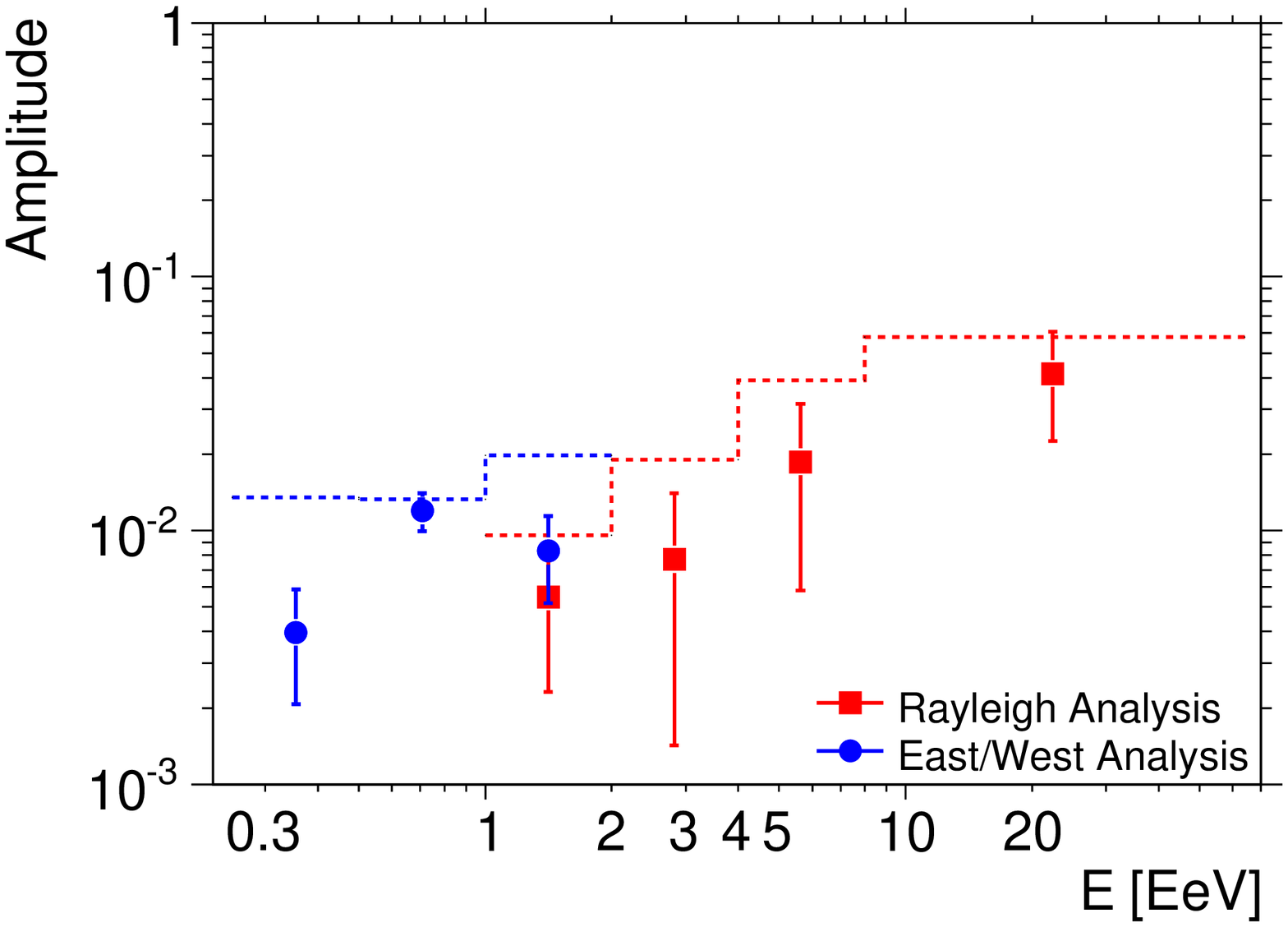}
  \includegraphics[width=10cm]{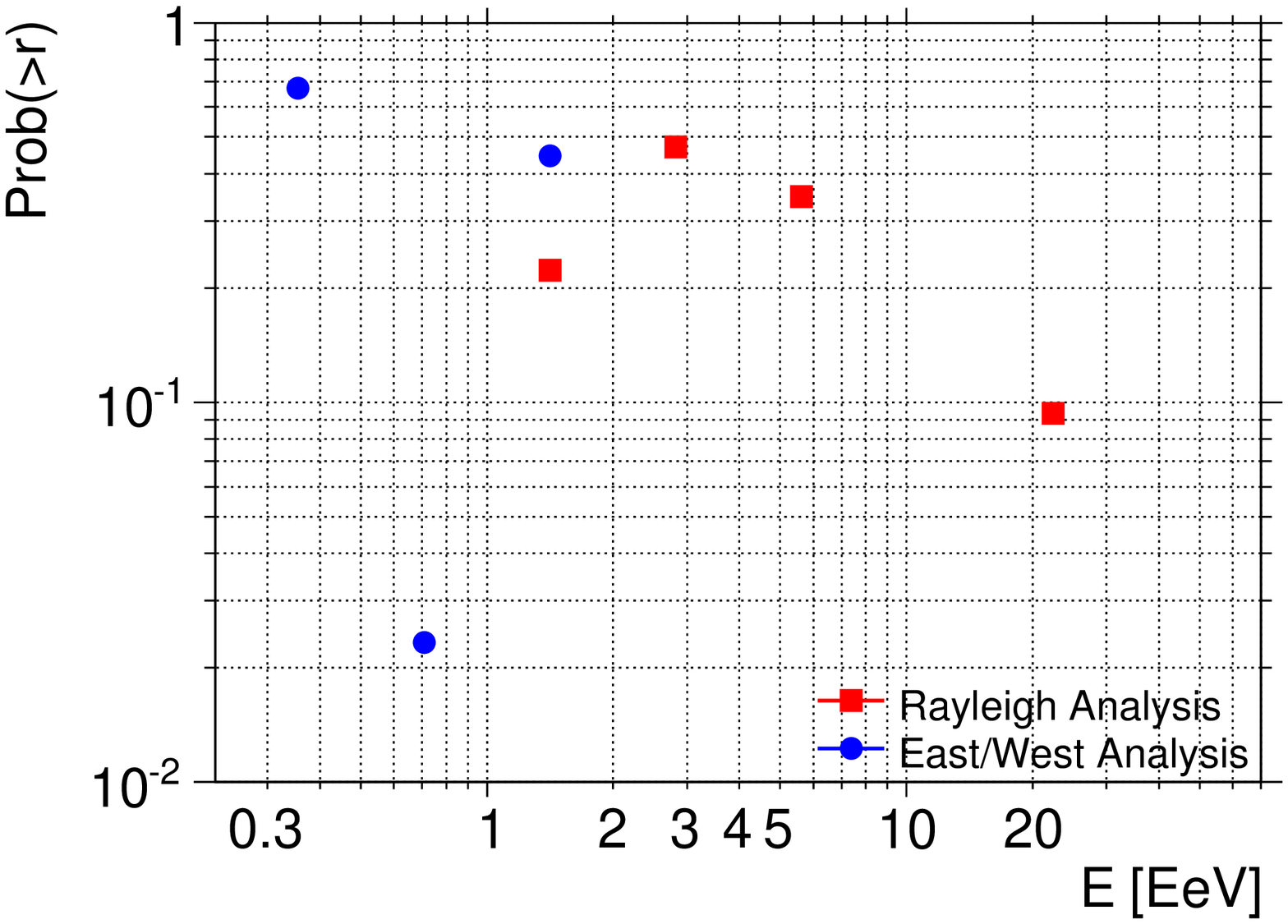}
  \caption{\small{Top~: Amplitude of the first harmonic as a function of energy. The dashed 
line indicates the 99\% $C.L.$ upper bound on the amplitudes that could result from fluctuations 
of an isotropic distribution. Bottom~: Corresponding probabilities to get at least the same 
amplitude from an underlying isotropic distribution.}}
  \label{amp_diff}
\end{figure}

\subsection{Results at the sidereal frequency in independent energy bins}

To perform first harmonic analyses as a function of energy, the choice of the size of the 
energy bins, although arbitrary, is important to avoid the dilution of a genuine 
signal with the background noise. In addition, the inclusion of intervals whose 
width is below the energy resolution or with too few data is most likely to weaken the 
sensitivity of the search for an energy-dependent anisotropy~\cite{alan}. To fulfill both 
requirements, the size of the energy intervals is chosen to be $\Delta\log_{10}(E)=0.3$ below 
8~EeV, so that it is larger than the energy resolution even at low energies. At higher
energies, to guarantee the determination of the amplitude measurement within an uncertainty 
$\sigma\simeq2\%$, all events ($\simeq 5,000$) with energies above 8~EeV are gathered  
in a single energy interval. 

The amplitude $r$ at the sidereal frequency as a function of the energy is shown in 
Fig.~\ref{amp_diff}, together with the corresponding probability $P(>r)$ to get a larger 
amplitude in each energy interval for a statistical fluctuation of isotropy. The dashed 
line indicates the 99\% $C.L.$ upper bound on the amplitudes that could result from 
fluctuations of an isotropic distribution.  It is apparent that there is no evidence of any 
significant signal over the whole energy range. A global statement refering to the probability with 
which the 6 observed amplitudes could have arisen from an underlying isotropic distribution can 
be made by comparing the measured value $K=\sum_{i=1}^6k_{0_i}$ (where the sum is over all 6 
independent energy intervals) with that expected from a random distribution for which 
$\left<K\right>=6$~\cite{edge}. The statistics of $2K$ under the hypothesis of an isotropic sky 
is a $\chi^2$ with $2\times6=12$ degrees of freedom. For our data, $2K=19.0$ and the associated 
probability for an equal or larger value arising from an isotropic sky is $\simeq 9$\%. 

\begin{figure}[!t]
  \centering					 
  \includegraphics[width=10cm]{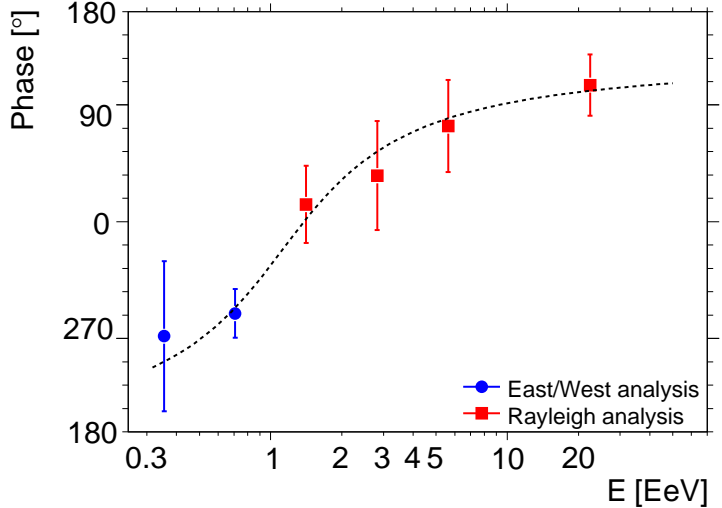}
  \caption{\small{Phase of the first harmonic as a function of energy. The dashed line,
  resulting from an empirical fit, is used in the likelihood ratio test (see text).}}
  \label{phase_diff}
\end{figure}

The phase $\varphi$ of the first harmonic is shown in Fig.~\ref{phase_diff} as a function 
of the energy. While the measurements of the amplitudes do not provide any evidence for 
anisotropy, we note that the measurements in adjacent energy intervals suggest a smooth 
transition between a common phase of $\simeq 270^\circ$ in the first two bins below 
$\simeq1$~EeV compatible with the right ascension of the Galactic Center 
$\alpha_{GC}\simeq268.4^\circ$, and another phase ($\alpha\simeq100^\circ$) above 
$\simeq5$~EeV. This is intriguing, as the phases are expected to be randomly distributed 
in case of independent samples whose parent distribution is isotropic. Knowing the p.d.f. 
of phase measurements drawn from an isotropic distribution, $p_0(\varphi)=(2\pi)^{-1}$, 
and drawn from a population of directions having a non-zero amplitude $r_0$ with a phase 
$\varphi_0$, $p_1(\varphi;r_0,\varphi_0)$~\cite{linsley-fh}, the likelihood functions of 
any of the hypotheses may be built as~:
\begin{equation}
\label{eqn:likelihoods}
L_0=\prod_{i=1}^{N_{bins}}p_0(\varphi_i), \hspace{1cm} L_1=\prod_{i=1}^{N_{bins}}p_1(\varphi_i;r_0,\varphi_0).
\end{equation}
Without any knowledge of the expected amplitudes $r_0(E)$ in each bin, the values considered 
in $L_1$ are the measurements performed in each energy interval. For the expected phases 
$\varphi_0(E)$ as a function of energy, we use an arctangent function adjusted on the data 
as illustrated by the dashed line in Fig.~\ref{phase_diff}. Since the smooth evolution 
of the phase distribution is potentially interesting but observed \textit{a posteriori}, 
we aim at testing the fraction of random samples whose behaviour in adjacent energy bins would 
show such a potential interest but with no reference to the specific values observed in the data.
To do so, we use the method of the likelihood ratio test, computing the $-2\ln(\lambda)$ statistic 
where $\lambda=L_0/L_1$. Using only $N_{bins}=6$, the asymptotic behaviour of the $-2\ln(\lambda)$ 
statistic is not reached. Hence, the p.d.f. of $-2\ln(\lambda)$ under the hypothesis of isotropy 
is built by repeating exactly the same procedure on a large number of isotropic samples~: in 
each sample, the arctangent parameters are left to be optimised, and the corresponding value 
of $-2\ln(\lambda)$ is calculated. In that way, any alignments, smooth evolutions or abrupt 
transitions of phases in random samples are captured and contribute to high values of 
the $-2\ln(\lambda)$ distribution. The probability that the hypothesis of isotropy better 
reproduces our phase measurements compared to the alternative hypothesis is then calculated 
by integrating the normalised distribution of $-2\ln(\lambda)$ above the value measured in the 
data. It is found to be $\simeq 2\cdot 10^{-3}$.

It is important to stress that no confidence level can be built from this report as we did not 
perform an \textit{a priori} search for a smooth transition in the phase measurements. To confirm 
the detection of a real transition using only the measurements of the phases with an independent 
data set, we need to collect $\simeq 1.8$ times the number of events analysed here to reach an 
efficiency of $\simeq 90\%$ to detect the transition at 99\% $C.L.$ (in case the observed effect 
is genuine). It is also worth noting that with a real underlying anisotropy, a consistency of the 
phase measurements in ordered energy intervals is expected with lower statistics than the detection 
of amplitudes significantly standing out of the background noise~\cite{edge,alan-linsley}. This 
behaviour was pointed out by Linsley, quoted in~\cite{edge}~: ``if the number of events available 
in an experiment is such that the RMS value of $r$ is equal to the true amplitude, then in a 
sequence of experiments $r$ will be significant (say $P(>r)<1\%$) in one experiment out of ten 
whereas the phase will be within 50$^\circ$ of the true phase in two experiments out of three.'' 
We have checked this result using Monte Carlo simulations. 

An apparent constancy of phase, even when the significances of the amplitudes are relatively
small, has been noted previously in surveys of measurements made in the range 
$10^{14}<E<10^{17}\,$eV~\cite{greisen,LW}. In~\cite{greisen} Greisen and his colleagues comment
that most experiments have been conducted at northern latitudes and therefore the reality of
the sidereal waves is not yet established. The present measurement is made with events coming 
largely from the southern hemisphere. 

\subsection{Additional cross-checks against systematic effects above 1$\,$EeV}
\label{addcrosschecks}

It is important to verify that the phase effect is not a manifestation of systematic effects, 
the amplitudes of which are at the level of the background noise. We provide hereafter additional 
studies above 1~EeV, where a few tests can cross-check results presented in Fig.~\ref{phase_diff}.

\begin{figure}[!t]
  \centering					 
  \includegraphics[width=7.5cm]{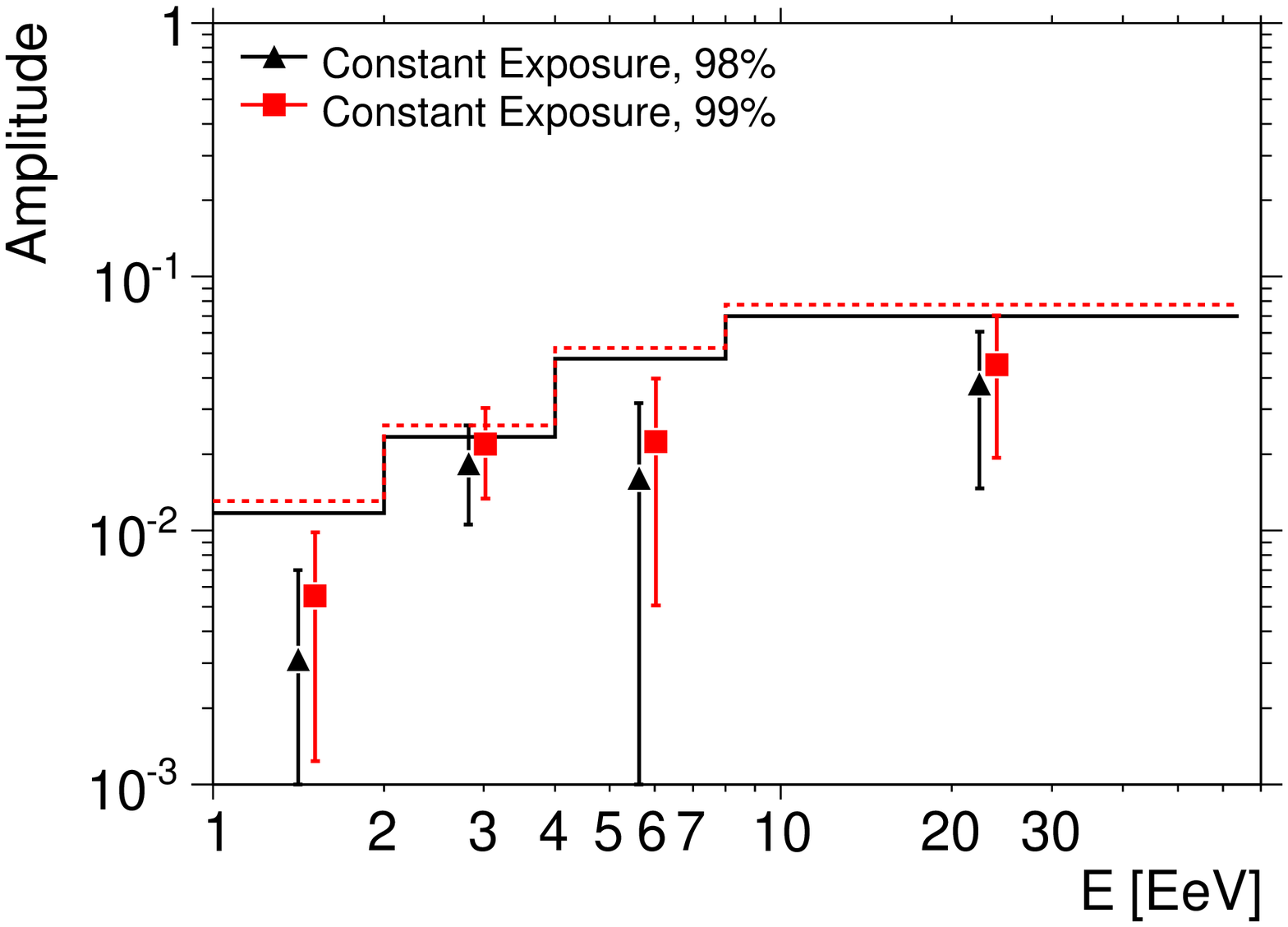}
  \includegraphics[width=7.5cm]{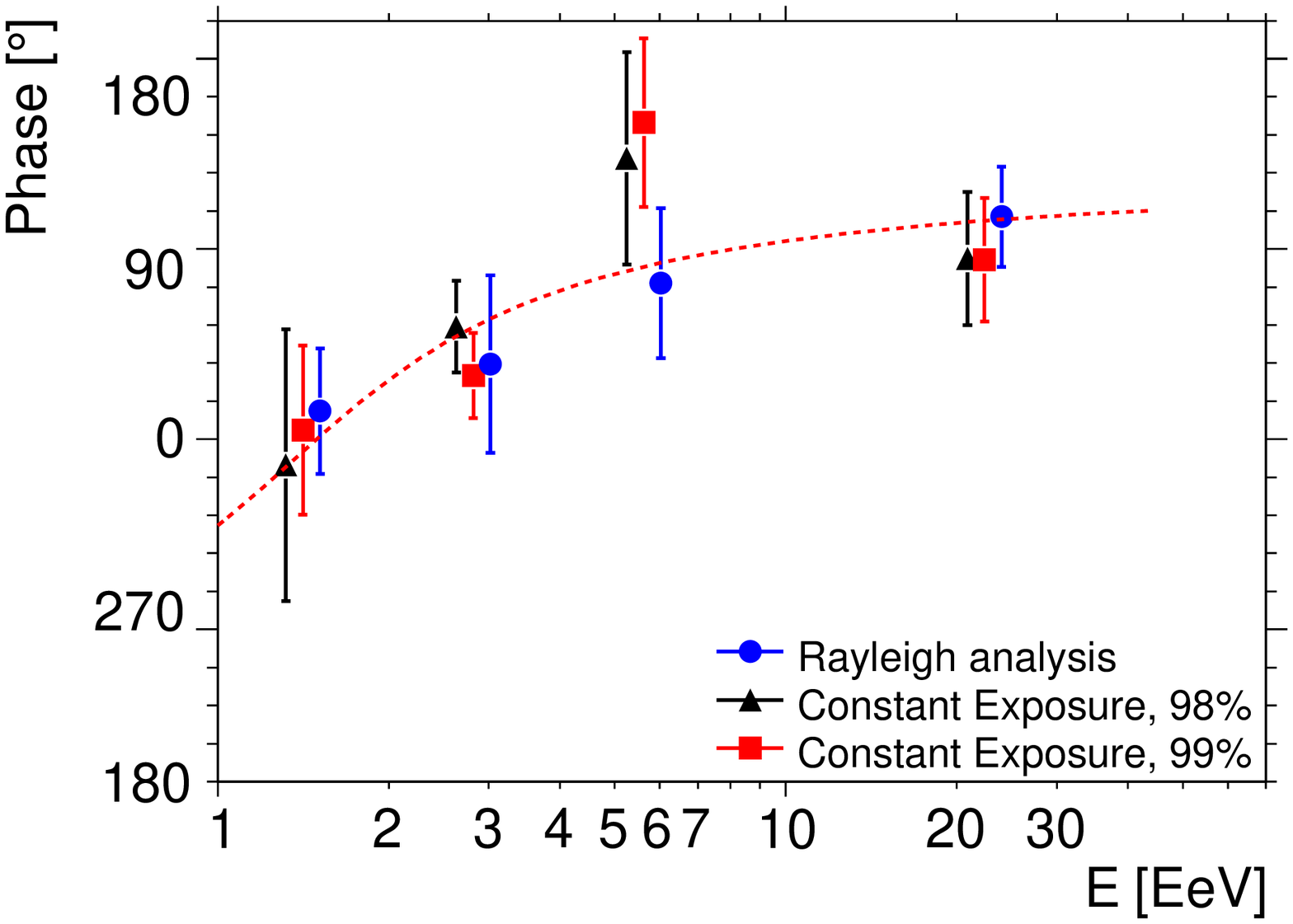}
  \caption{\small{Rayleigh analysis above 1~EeV using a reduced data set built to get at a constant 
exposure in right-ascension, using on-times of 98\% (triangles) and 99\% (squares) (see text). 
Left~: analysis of the amplitude. The full (dashed) line indicates the 99\% $C.L.$ upper bound on 
the amplitudes that could result from fluctuations of an isotropic distribution when using on-times 
of 98\% (99\%). Right~: analysis of the phase, the dashed line being the same as the one plotted 
in Fig.~\ref{phase_diff}. Results of Fig.~\ref{phase_diff} are also shown with circles.}}
  \label{syst_cstexpo} 
\end{figure}

The first cross-check is provided by applying the Rayleigh analysis on a reduced data set built 
in such a way that its corresponding exposure in right ascension is uniform. This can be achieved 
by selecting for each sidereal day only events triggering an unitary cell whose on-time was almost 
100\% over the whole sidereal day. To keep a reasonably large data set, we present here the results 
obtained for on-times of 98\% and 99\%. This allows us to use respectively $\simeq77$\% 
and $\simeq63$\% of the cumulative data set without applying any correction to account for a 
non-constant exposure. The results are shown in Fig.~\ref{syst_cstexpo} when considering on-time 
of 98\% (triangles) and 99\% (squares). Even if more noisy due to the reduction of the statistics 
with respect to the Rayleigh analysis applied on the cumulative data set, they are consistent 
with the weighted Rayleigh analysis and support that results presented in Fig.~\ref{phase_diff} 
are not dominated by any residual systematics induced by the non-uniform exposure. 

\begin{figure}[!t]
  \centering					 
  \includegraphics[width=7.5cm]{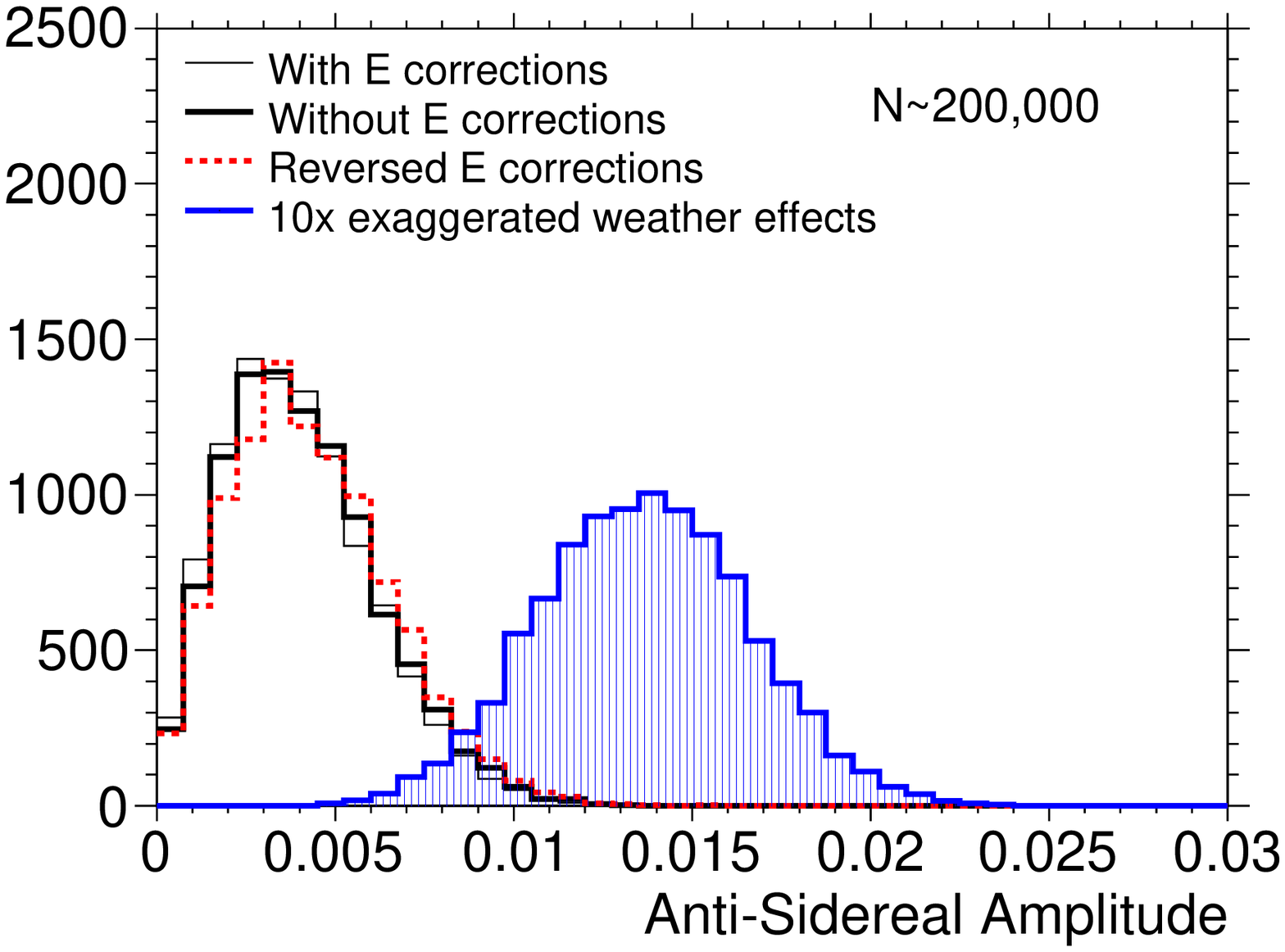}
  \includegraphics[width=7.5cm]{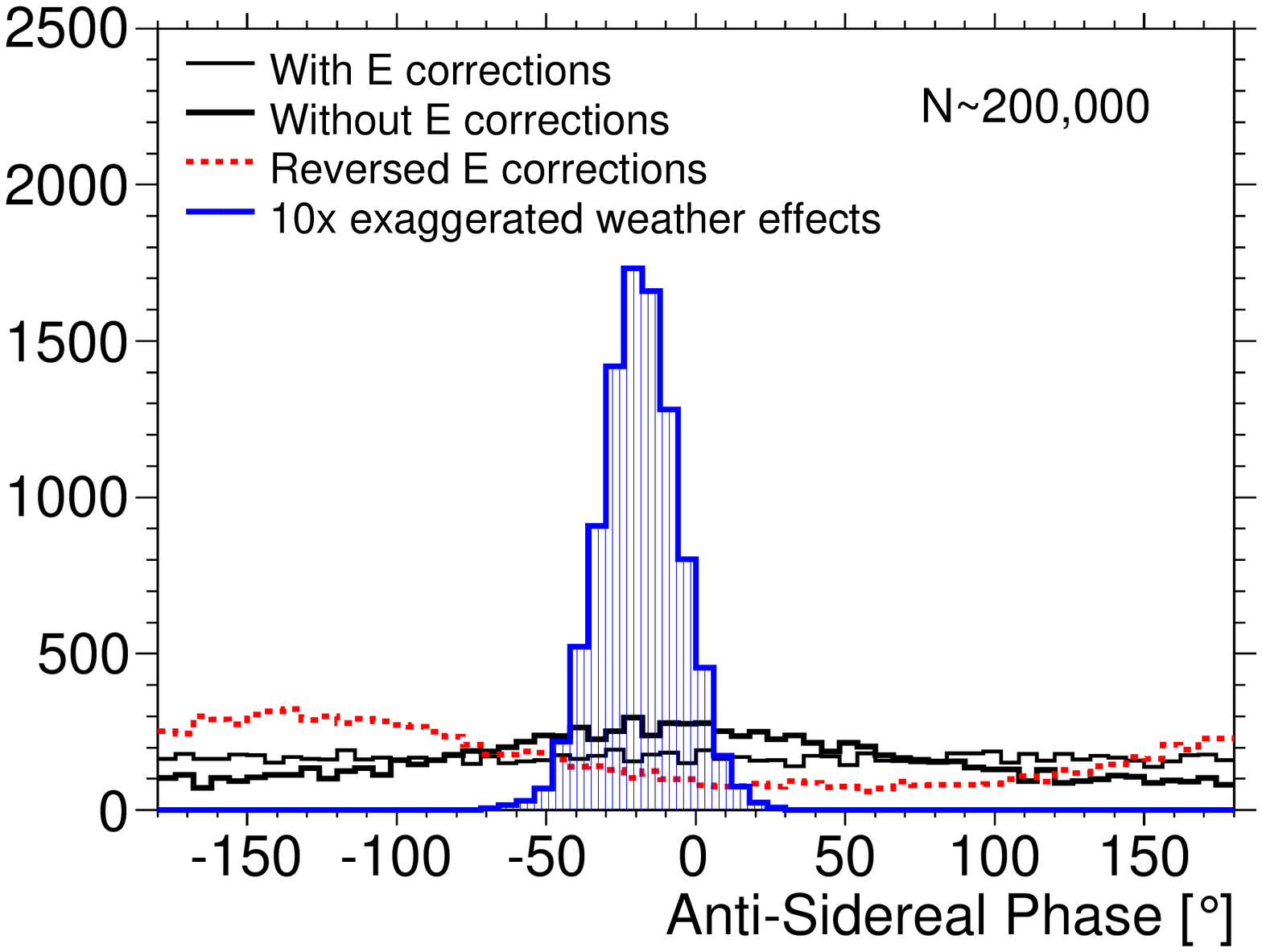}
  \includegraphics[width=7.5cm]{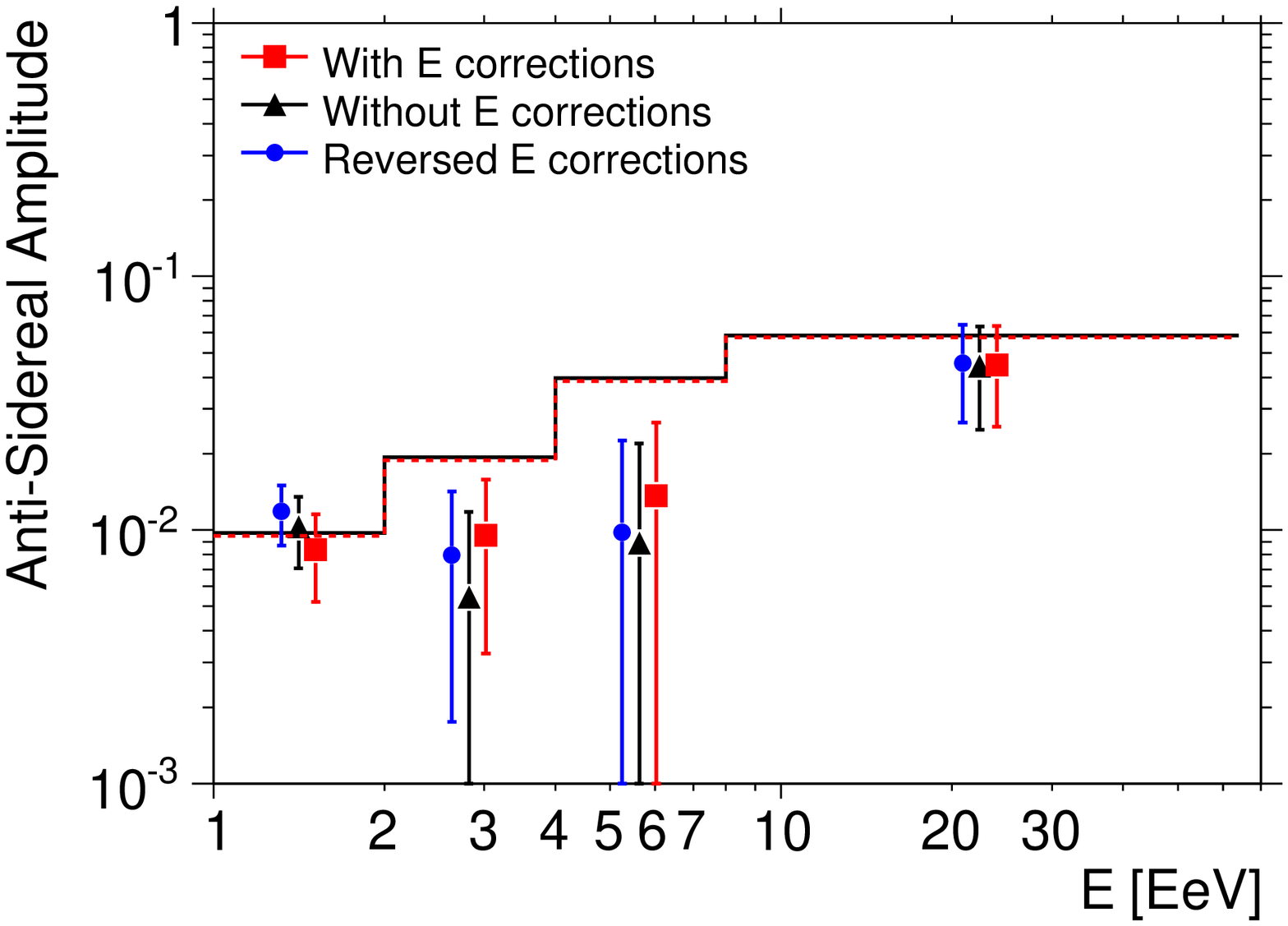}
  \includegraphics[width=7.5cm]{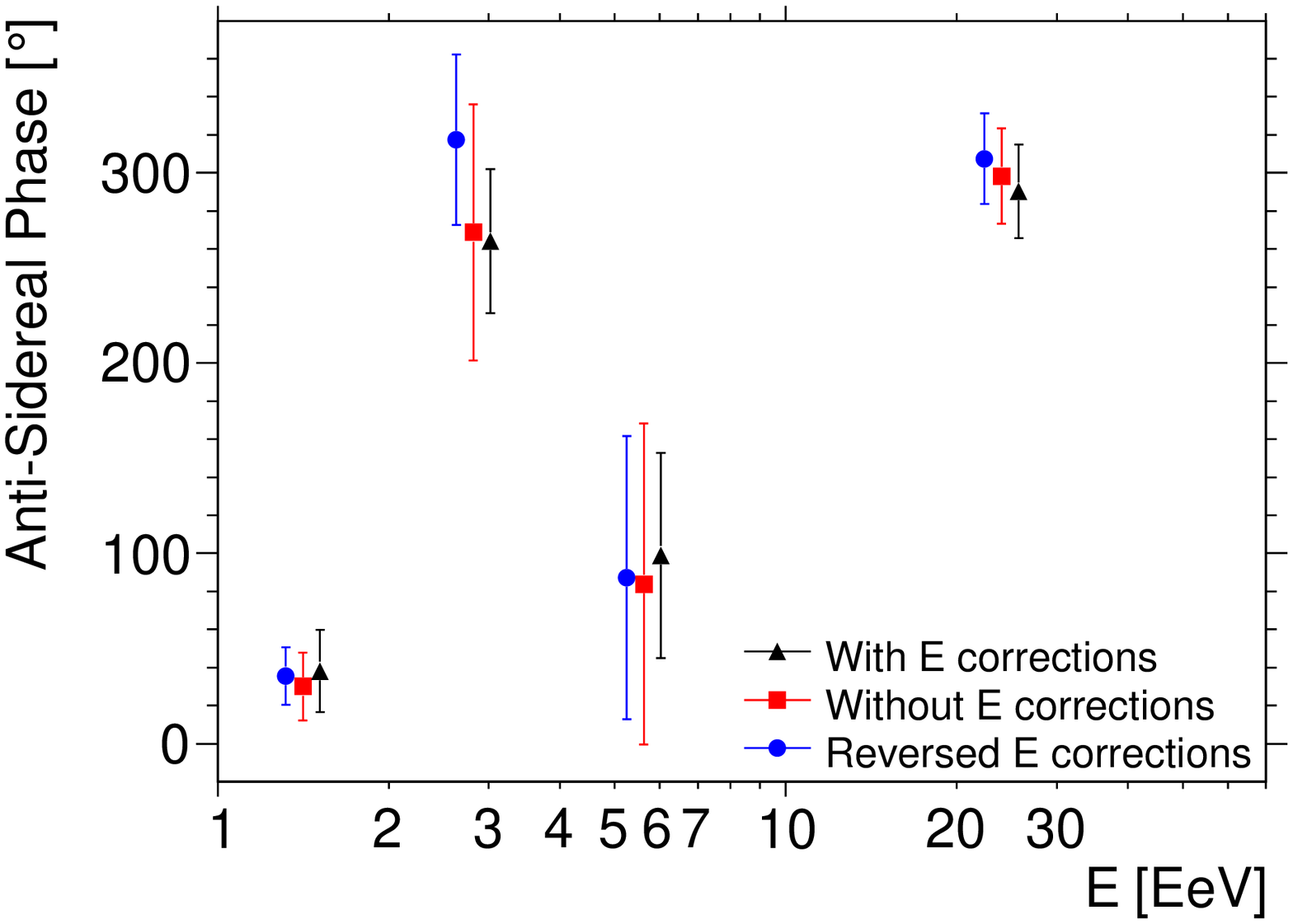}
  \caption{\small{Top~:Anti-sidereal analysis of mock samples built such that a spurious 
anti-sidereal amplitude stands out from the noise through the sideband mechanism between 
1 and 2~EeV (see text). Thin histograms~: standard Rayleigh analysis of the amplitudes 
(left panel) and phases (right panel) for isotropic samples. Thick histograms~: same for 
samples biased by the energy variations induced by the atmospheric changes. Dashed histograms~: 
same by \emph{reversing} the energy corrections. Filled histograms~: same by \emph{amplifying} by 
10 the energy variations. Bottom~: Rayleigh analysis of the anti-sidereal amplitude (left) and 
phase (right) of the real data sample without applying the energy corrections (squares), by 
applying those corrections (triangles), and by reversing the energy corrections (circles).}}
  \label{syst_antisid} 
\end{figure}

From the Fourier analysis presented in section~\ref{sec:solar-antisid}, we have stressed
the decoupling between the solar frequency and both the sidereal and anti-sidereal ones
thanks to the frequency resolution reached after 6 years of data taking. However, as the 
amplitude of an eventual sideband effect is \emph{proportional} to the solar amplitude
~\cite{far-sto}, it remains important to estimate the impact of an eventual sideband 
effect persisting even after the energy corrections. To probe the magnitude of this sideband 
effect, we use 10,000 mock data sets generated from the real data set (with energies 
corrected for weather effects) by randomising the arrival times but meanwhile keeping
both the zenith and the azimuth angles of each original event. This procedure guarantees 
the production of isotropic samples drawn from a uniform exposure with the same detection 
efficiency conditions than the real data. The results of the Rayleigh analysis applied to 
each mock sample between 1 and 2~EeV at the \emph{anti-sidereal} frequency are shown by the thin 
histograms in top panels of Fig.~\ref{syst_antisid}, displaying Rayleigh distributions
for the amplitude measurements and uniform distributions for the phase measurements. Then, 
after introducing in to each sample the temporal variations of the energies induced by the atmospheric 
changes according to Eqn.~\ref{sweather}, it can be seen on the same graph (thick
histograms) that the amplitude measurements are almost undistinguishable with respect to 
the reference ones, while the phase measurements start to show to a small extent a 
preferential direction. The same conclusions hold when \emph{reversing} the energy corrections 
(dashed histograms), but resulting in a phase shift of $\simeq 180^\circ$. Finally, 
the filled histograms are obtained by \emph{amplifying} by 10 the energy variations induced by 
the atmospheric changes. In this latter case, the large increase of the solar amplitude 
induces a clear signal at the anti-sidereal frequency through the sideband mechanism, 
as evidenced by the distributions of both the amplitudes and the phases. The sharp maximum 
of the phase distribution points towards the spurious direction, while the amplitude 
distribution follows a non-centered Rayleigh distribution with parameter $\simeq 1.4\times 10^{-2}$. 
The spurious mean amplitude stands out from the noise ($\sim\sqrt{\pi/200,000}\simeq 4\times 10^{-3}$) 
sufficiently to allow us to estimate empirically the original effect to be ten times 
smaller, at the level of $\simeq 1.4\times 10^{-3}$. This will impact the analyses only in a marginal 
way even if we had not performed the energy corrections, or if we had over-corrected the energies.
The energy corrections necessarily reduce even more the size of the sideband amplitude, well 
below  $\simeq 10^{-3}$. Hence, only small changes are expected in the anti-sidereal phase 
measurements on the real data when applying (or not) or reversing the energy corrections. This is 
found to be the case, as illustrated in the bottom panels of Fig.~\ref{syst_antisid}. In addition, 
this phase distribution (bottom right panel of Fig.~\ref{syst_antisid}) does not show any 
particular structure aligned with the spurious direction. This second cross-check support the 
hypothesis that the phase measurements presented in Fig.~\ref{phase_diff} are not dominated by 
any residual systematics induced by the sideband mechanism.  

\begin{figure}[!t]
  \centering					 
  \includegraphics[width=7.5cm]{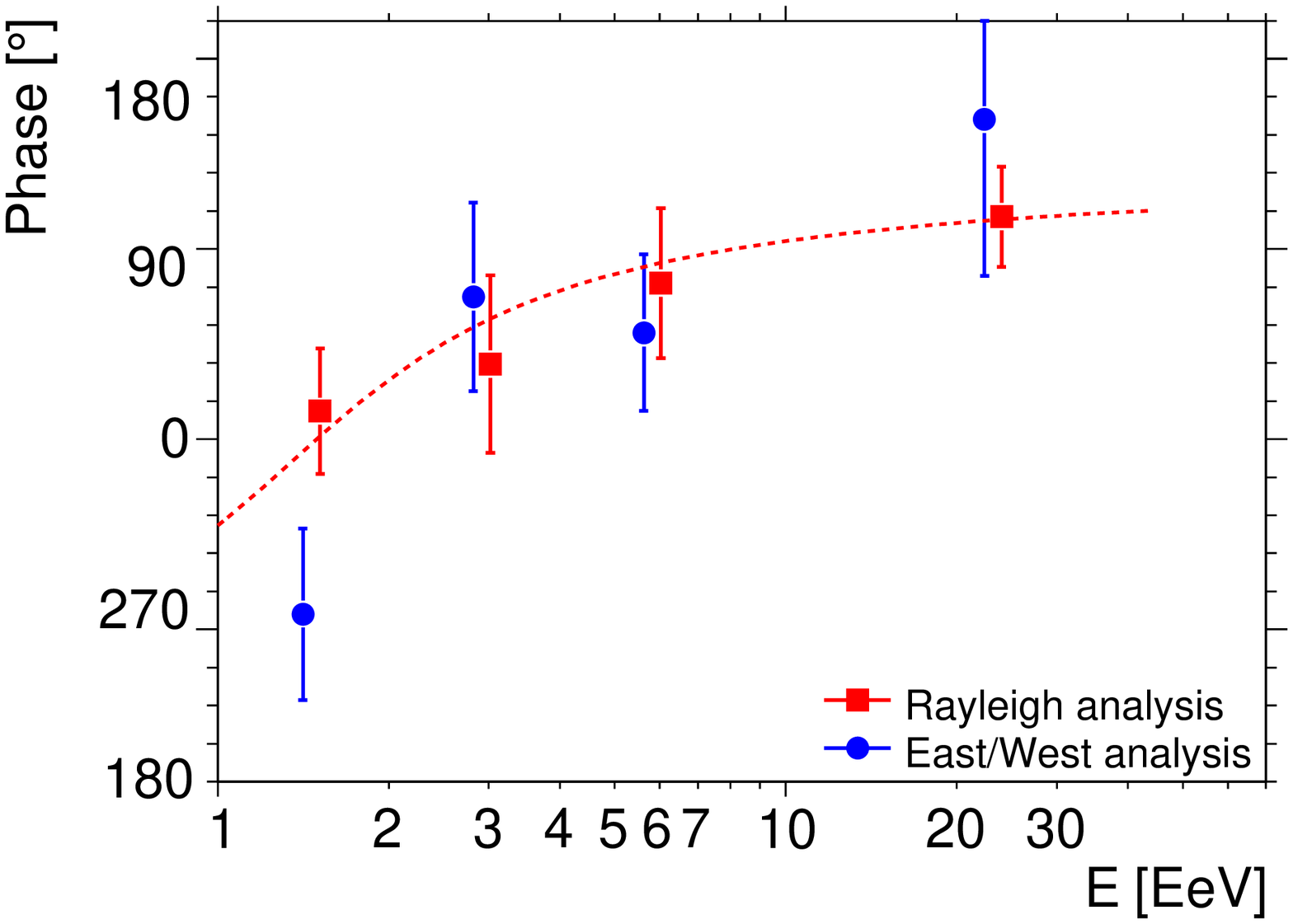}
  \includegraphics[width=7.5cm]{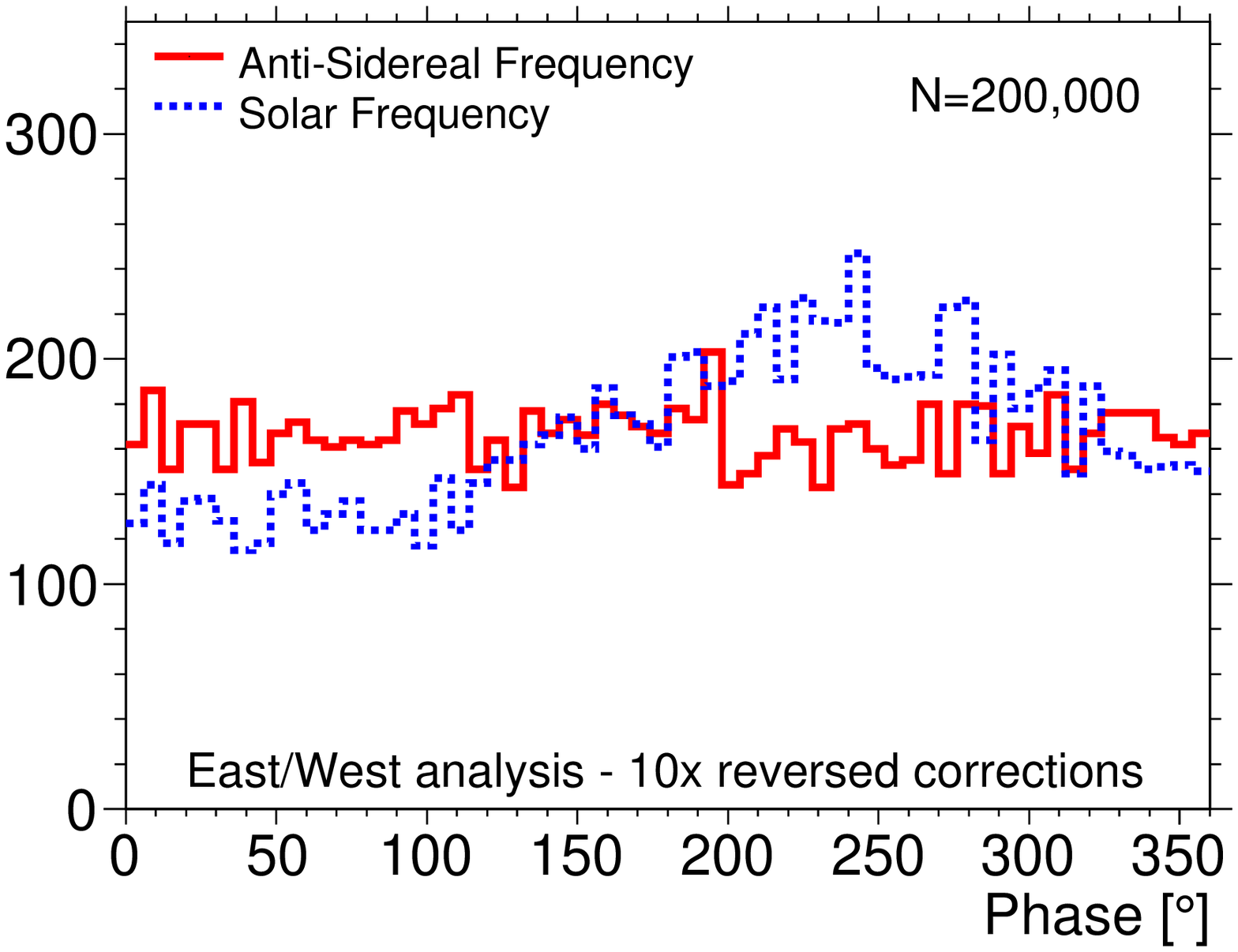}
  \caption{\small{Left~: East/West analysis at the sidereal frequency above 1~EeV. Right~: 
East/West analyses at both the solar (dashed) and the anti-sidereal (thick) frequencies of mock 
samples built by amplifying by 10 the energy variations induced by weather effects.}}
  \label{ew_syst}
\end{figure}

In the left panel of Fig.~\ref{ew_syst} we show the results of the East/West analysis (circles). 
They provide further support of the previous analyses (squares). As previously explained, 
this method relies on the high symmetry between the Eastern and Western sectors. 
However, the array being slightly tilted ($\simeq 0.2^\circ$), this symmetry is slightly broken,
resulting in a small shift between the Eastward and Westward counting rates. As this shift is
independent of time, it does not impact itself in the estimate of the first harmonic. However, 
it is worth examining the effect of the \emph{combination} of the tilted array together with 
the spurious modulations induced by weather effects, as this combination may mimic a real 
East/West first harmonic modulation at the solar frequency. The size of such an effect can be probed 
by analysing the mock data sets built by amplifying by 10 the energy variations induced by the 
atmospheric changes. The results obtained between 1 and 2~EeV at both the solar and the 
anti-sidereal frequencies are shown in the right panel of Fig.~\ref{ew_syst}~: while the phase 
distribution starts to show a preferential direction at the solar frequency (dashed histogram), 
the same distribution is still uniform at the anti-sidereal one (thick histogram). Hence, it is 
safe to conclude that the results obtained at the sidereal frequency by means of the East/West 
method are not affected by any systematics. 
 
\subsection{Results at the sidereal frequency in cumulative energy bins}

\begin{figure}[!t]
  \centering					 
  \includegraphics[width=10cm]{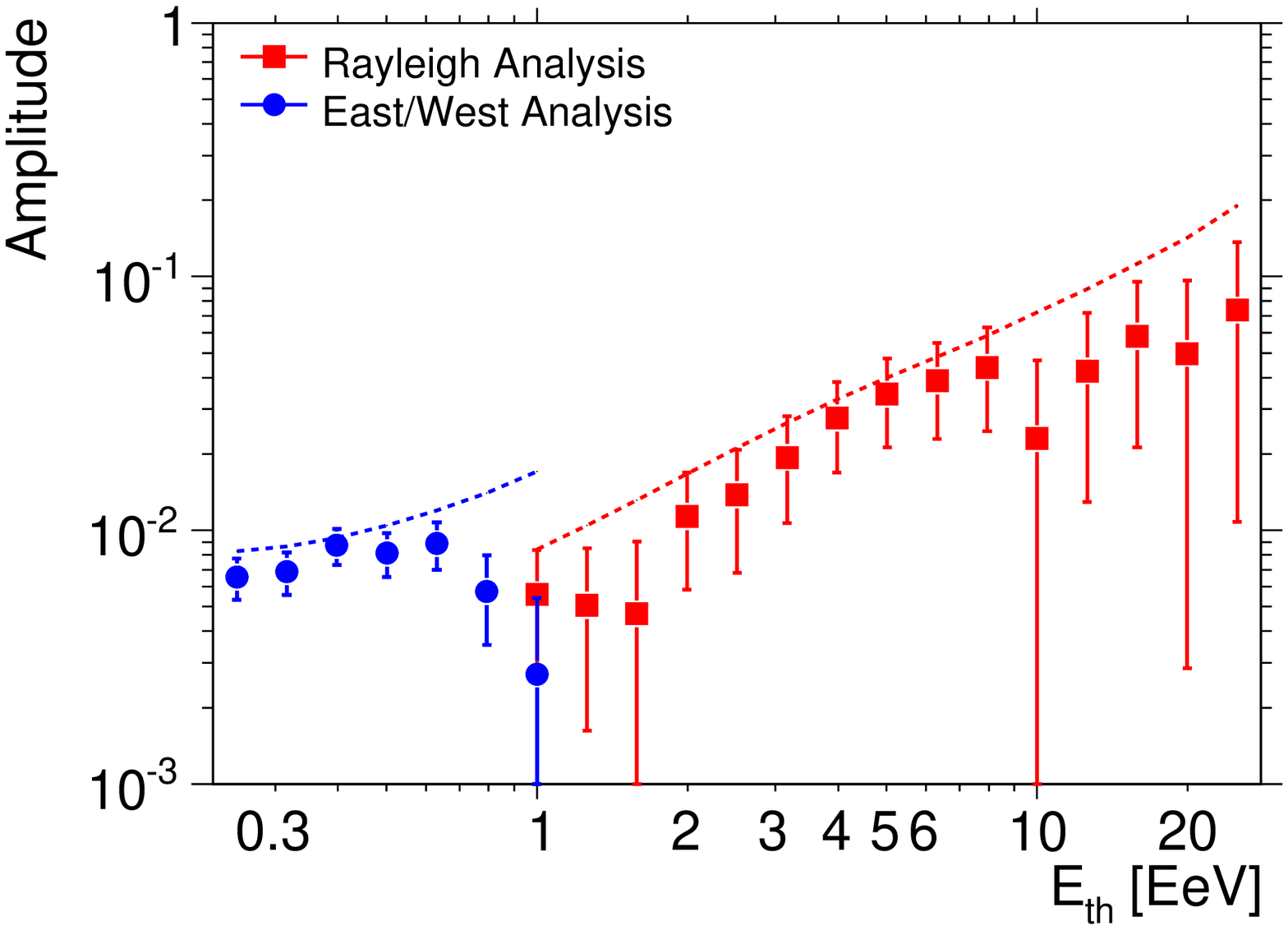}
  \includegraphics[width=10cm]{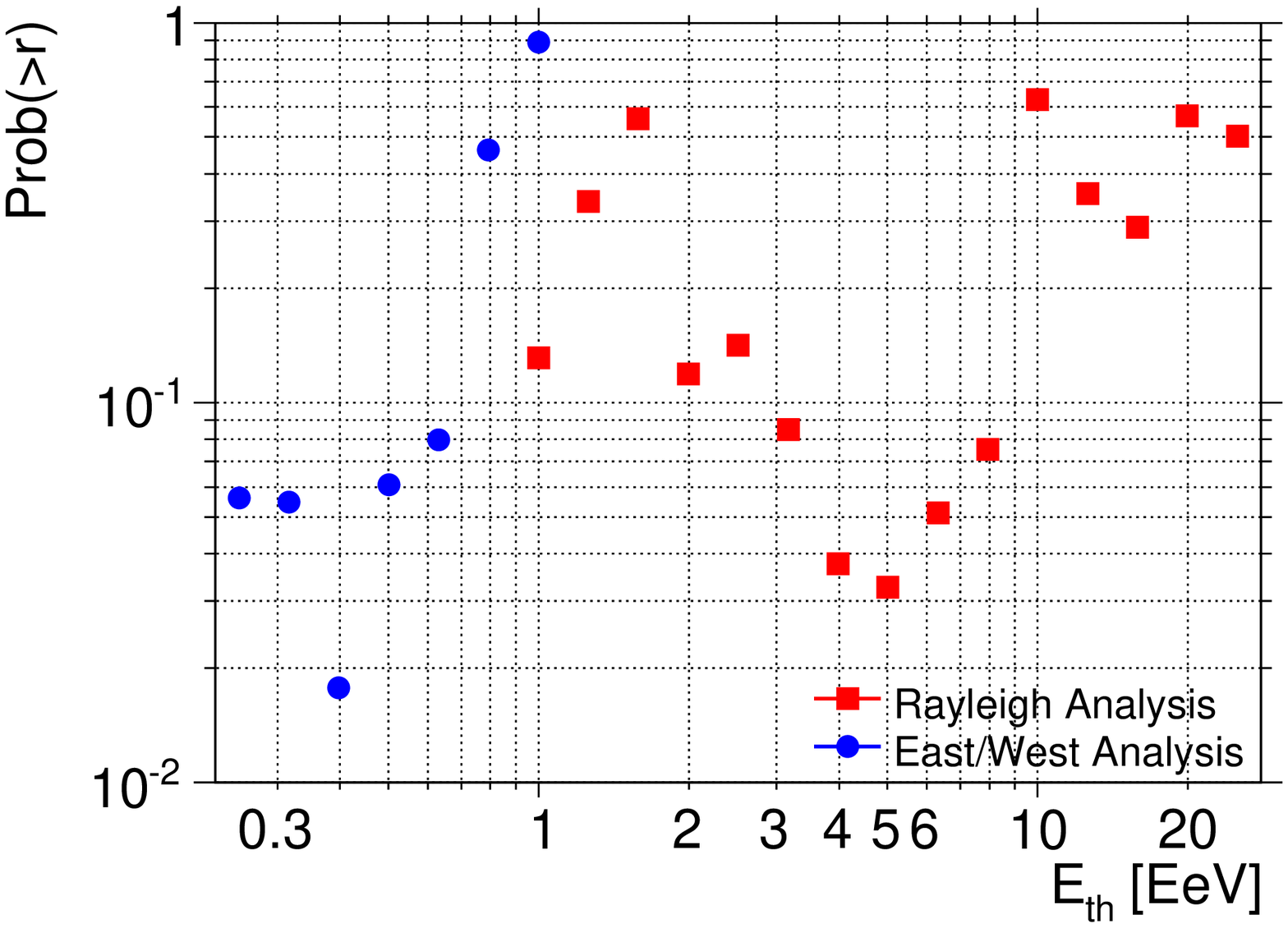}
  \includegraphics[width=10cm]{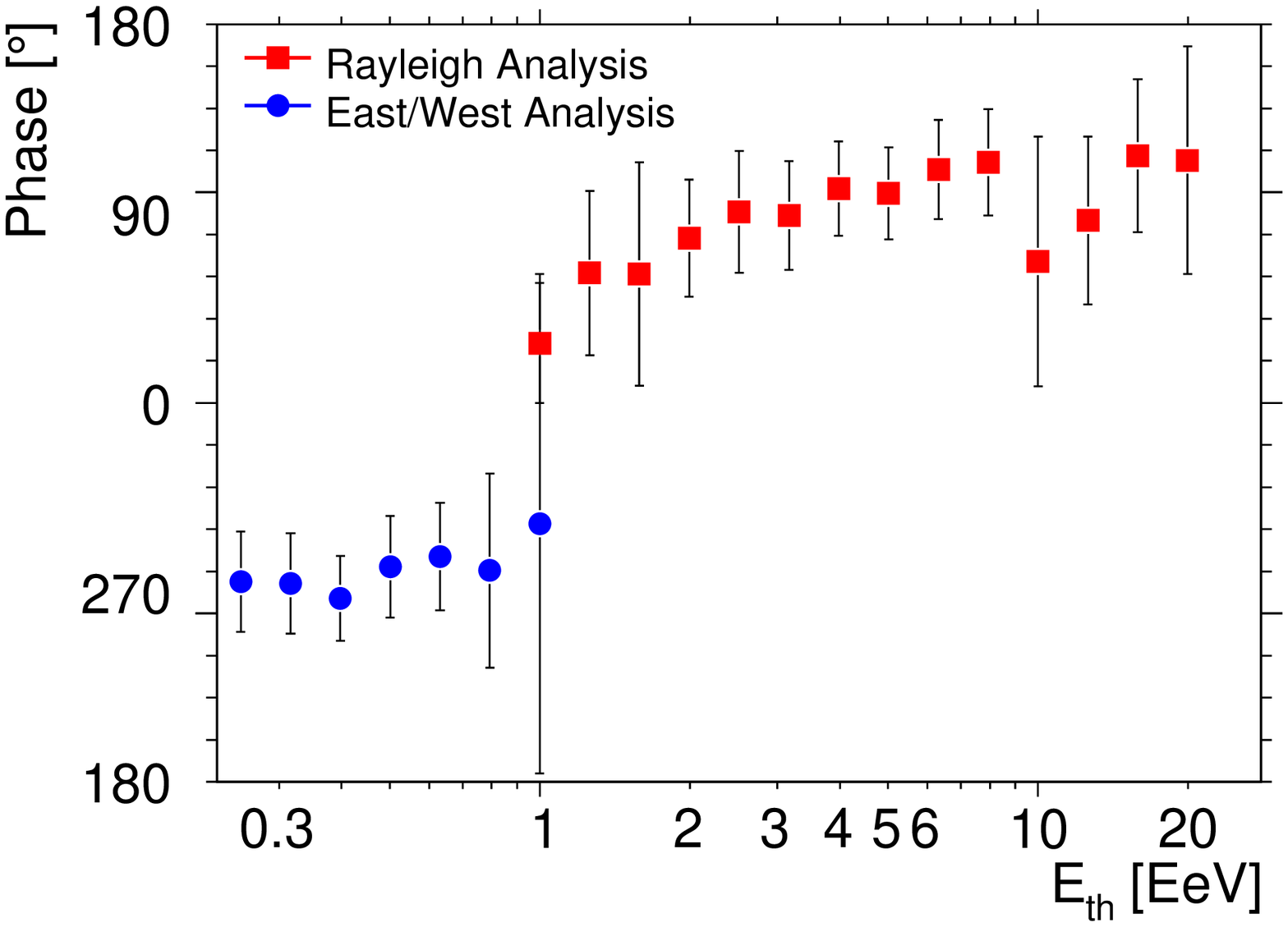}
  \caption{\small{Same as Fig.~\ref{amp_diff} and Fig.~\ref{phase_diff}, but as a function of 
energy thresholds.}}
  \label{amp_thres}
\end{figure}

Performing the same analysis in terms of energy thresholds may be convenient for optimizing the
detection of an eventual genuine signal spread over a large energy range, avoiding the arbitrary 
choice of a bin size $\Delta\log_{10}(E)$. The bins are however strongly correlated, preventing 
a straightforward interpretation of the evolution of the points with energy. The results on the 
amplitudes are shown in Fig.~\ref{amp_thres}. They do not provide any further evidence in favor 
of a significant amplitude. 

\section{Upper limits and discussion}
\label{sec:discussion}

\begin{table*}[t]
\begin{center}
\begin{tabular}{c|c|c|c|c|c|c}
$\Delta E$ & $N$ &  $r_{\mathrm{sidereal}}$[\%] & $P(>r_{\mathrm{sidereal}})[\%]$ & $\varphi$ [$^\circ$]& $\Delta \varphi$ [$^\circ$] & $d_\perp^{UL}$ [\%]\\
\hline
\hline
0.25 - 0.5 & 553639 & 0.4 & 67 & 262 & 64 & 1.3\\
0.5 - 1 & 488587 & 1.2 & 2 & 281 & 20 & 1.7\\
1 - 2 & 199926 & 0.5 & 22 & 15 & 33& 1.4\\
2 - 4 & 50605 & 0.8 & 47 & 39 & 46 & 2.3\\
4 - 8 & 12097 & 1.8 & 35 & 82 & 39 & 5.5\\
$>$8 & 5486 & 4.1 & 9 & 117 & 27 & 9.9
\end{tabular}
\caption{\small{Results of first harmonic analyses in different energy intervals, using the East/West 
analysis below 1~EeV and the Rayleigh analysis above 1~EeV.}}
\label{tab2}
\end{center}
\end{table*}%

\begin{figure}[t]
  \centering	
  \includegraphics[width=13cm]{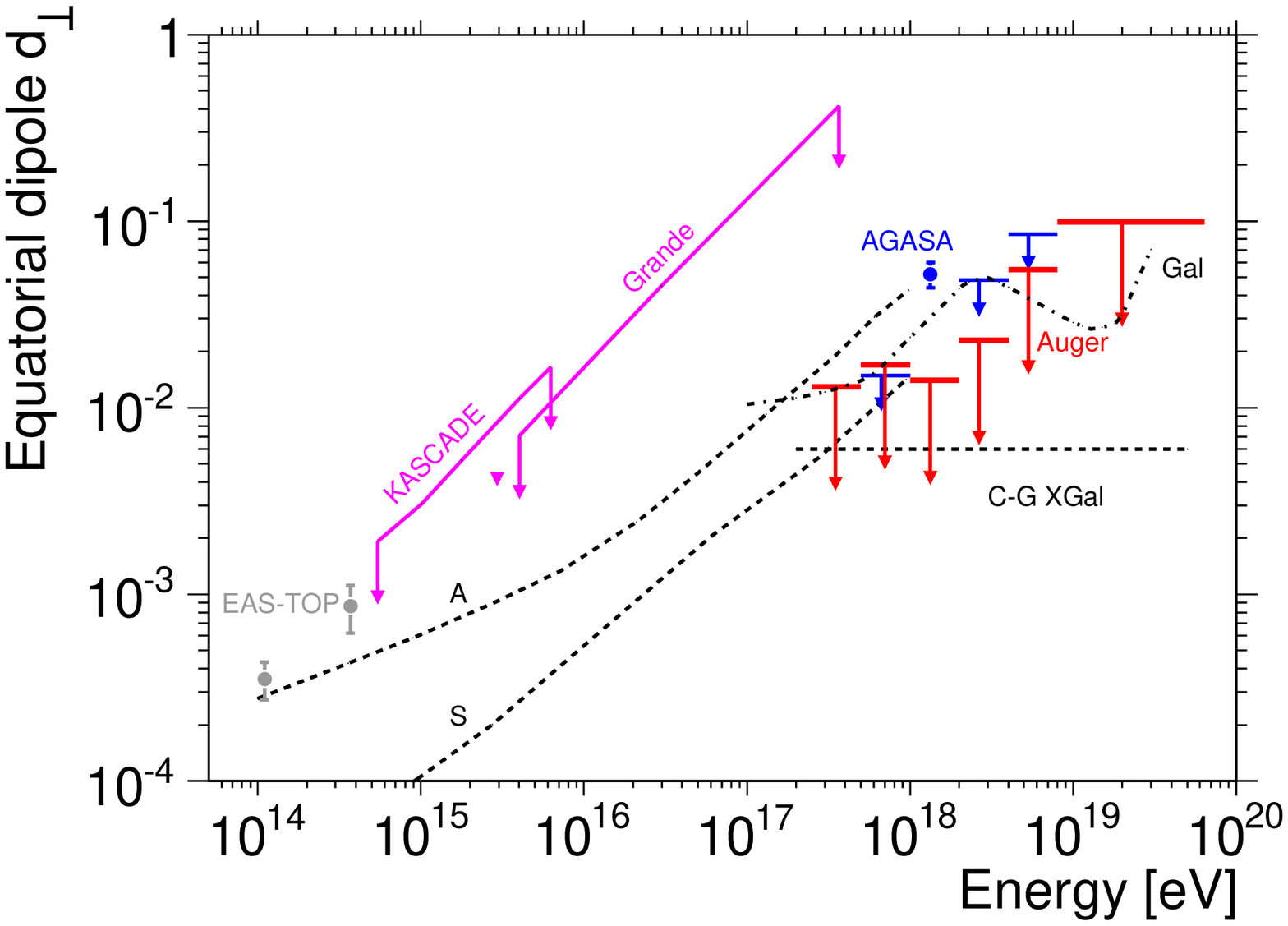}
  \caption{\small{Upper limits on the anisotropy amplitude of first harmonic as a function of 
energy from this analysis. Results from EAS-TOP, AGASA, KASCADE and KASCADE-Grande experiments 
are displayed too. An analysis of the KASCADE-Grande data with the East/West method delivers 
an additional limit for $3\,10^{15}\,$eV. 
Also shown are the predictions up to 1~EeV from two different galactic magnetic field models with 
different symmetries ($A$ and $S$), the predictions for a purely galactic origin of UHECRs up to 
a few tens of $10^{19}\,$eV ($Gal$), and the expectations from the Compton-Getting effect for an 
extragalactic component isotropic in the CMB rest frame ($C$-$G\,Xgal$).}}
  \label{UL}
\end{figure}

From the analyses reported in the previous Section, upper limits on amplitudes at 99\% $C.L.$ can 
be derived according to the distribution drawn from a population characterised by an anisotropy of 
unknown amplitude and phase as derived by Linsley~\cite{linsley-fh}~:
\begin{equation}
\label{eqn:ul}
\sqrt{\frac{2}{\pi}}\frac{1}{I_0(r^2/4\sigma^2)}\int_0^{r_{UL}} \frac{\mathrm{d}s}{\sigma}\,I_0\bigg(\frac{rs}{\sigma^2}\bigg)\,\exp{\bigg(-\frac{s^2+r^2/2}{2\sigma^2}\bigg)}=C.L.,
\end{equation}
where $I_0$ is the modified Bessel function of the first kind with order 0, and $\sigma=\sqrt{2/\mathcal{N}}$ 
in case of the Rayleigh analysis, and $\sigma_{EW}=(\pi\left<\cos{\delta}\right>/2\left<\sin{\theta}\right>)\times\sqrt{2/N}$
in case of the East/West analysis. 

As discussed in the Appendix, the Rayleigh amplitude measured by an observatory 
depends on its latitude and on the range of zenith angles considered. 
The measured amplitude can be related to a real equatorial dipole component
$d_\perp$ by $d_\perp \simeq r/\langle \cos \delta\rangle$. This is the physical
quantity of interest to compare results from different experiments and from model 
predictions. The upper limits on $d_\perp$ are given in Tab.~\ref{tab2} and shown 
in Fig.\ref{UL}, together with previous results from EAS-TOP~\cite{eastop}, 
KASCADE~\cite{kascade}, KASCADE-Grande~\cite{grande} and AGASA~\cite{agasa}, and 
with some predictions for the anisotropies arising from models of both galactic and 
extragalactic UHECR origin. The results obtained in this study are not consistent 
with the $\simeq4$\% anisotropy reported by AGASA in the energy range $1<E/\mathrm{EeV}<2$. 

If the galactic/extragalactic transition occurs at the ankle energy~\cite{linsley-ankle}, 
UHECRs at 1$\,$EeV are predominantly of galactic origin and their escape from the galaxy 
by diffusion and drift motions are expected to induce a modulation in this energy range. 
These predictions depend on the assumed galactic magnetic field model as well as 
on the source distribution and the composition of the UHECRs\footnote{The dependence of 
the detection efficiency on the primary mass below 3$\,$EeV could affect the details of a direct
comparison with a model based on a mixed composition.}. Two alternative models are displayed 
in Fig. \ref{UL}, corresponding to different geometries of the halo magnetic fields~\cite{roulet1}. 
The bounds reported here already exclude the particular model with an antisymmetric halo 
magnetic field ($A$) and are starting to become sensitive to the predictions of the model 
with a symmetric field ($S$). We note that those models assume a predominantly heavy 
composition galactic component at EeV energies, while scenarios in which galactic protons 
dominate at those energies would typically predict anisotropies larger than the bounds 
obtained in Fig.~\ref{UL}. Maintaining the amplitudes of such anisotropies within our 
bounds necessarily translates into constraints upon the description of the halo magnetic 
fields and/or the spatial source distribution. This is particularly interesting in the 
view of our composition measurements at those energies compatible with a light 
composition~\cite{auger-er}. Aternatively to a leaky galaxy model, there is still the 
possibility that a large scale magnetic field retains all particles in the
galaxy~\cite{peters,jokipii}. If the structure of the magnetic fields in the halo is
such that the turbulent component predominates over the regular one, purely 
diffusion motions may confine light elements of galactic origin up to $\simeq 1\,$EeV 
and may induce an ankle feature due to the longer confinement of heavier elements at
higher energies~\cite{calvez}. Typical signatures of such a scenario in terms of large scale 
anisotropies are also shown in Fig.~\ref{UL} (dotted line)~: the corresponding amplitudes are 
challenged by our current sensitivity. 

On the other hand, if the transition is taking place at lower energies around the 
second knee at $\simeq5\times10^{17}\,$eV~\cite{berezinsky}, UHECRs above 1~EeV are 
dominantly of extragalactic origin and their large scale distribution could be influenced 
by the relative motion of the observer with respect to the frame of the sources. If the 
frame in which the UHECR distribution is isotropic coincides with the CMB rest frame, 
a small anisotropy is expected due to the Compton-Getting effect. Neglecting the effects 
of the galactic magnetic field, this anisotropy would be a dipolar pattern pointing in 
the direction $\alpha\simeq 168^\circ$ with an amplitude of about 0.6\%~\cite{kachelriess}. 
On the contrary, when accounting for the galactic magnetic field, this dipolar anisotropy 
is expected to also affect higher order multipoles~\cite{silvia}. These amplitudes are close 
to the upper limits set in this analysis, and the statistics required to detect an amplitude 
of 0.6\% at 99\% $C.L.$ is $\simeq 3$ times the present one. 

Continued scrutiny of the large scale distribution of arrival directions of UHECRs as a 
function of energy with the increased statistics provided by the Pierre Auger Observatory,
above a few times $10^{17}\,$eV, will help to discriminate between a predominantly galactic 
or extragalactic origin of UHECRs as a function of the energy, and so benefit the search
for the galactic/extragalactic transition. Future work will profit from the lower energy
threshold that is now available at the Pierre Auger Observatory~\cite{infill}.

\section*{}

This article is dedicated to \emph{Gianni Navarra}, who has been deeply involved in this study for 
many years and who has inspired several of the analyses described in this paper. His legacy lives
on.

\section*{Appendix}

The first harmonic amplitude of the distribution in right ascension of the detected cosmic rays can be 
directly related to the amplitude $d$ of a dipolar distribution of the cosmic ray flux of the form 
$J(\alpha,\delta) = J_0 (1+ d \ \hat d \cdot \hat u)$, where $\hat u$ and $\hat d$ denote respectively the 
unit vector in the direction of an arrival direction and in the direction of the dipole. Eqn.~\ref{eqn:fh} 
and Eqn.~\ref{eqn:fh2} can be used to express $a$, $b$ and $\mathcal{N}$ as~:
\begin{eqnarray}
a&=& \frac{2}{\mathcal{N}} \int_{\delta_{min}}^{\delta_{max}} d\delta \int_0^{2\pi}
d\alpha \cos \delta \ J(\alpha,\delta) \ \omega(\delta) \cos \alpha,\\  
b&=& \frac{2}{\mathcal{N}} \int_{\delta_{min}}^{\delta_{max}} d\delta \int_0^{2\pi}
d\alpha \cos \delta \ J(\alpha,\delta) \ \omega(\delta) \sin \alpha, \nonumber \\
\mathcal{N}&=& \int_{\delta_{min}}^{\delta_{max}} d\delta \int_0^{2\pi}
d\alpha \cos \delta \ J(\alpha,\delta) \ \omega(\delta),\nonumber
\end{eqnarray}
where we have here neglected in the exposure $\omega$ the small dependence on right 
ascension. Writing the angular dependence in 
$J(\alpha,\delta)$ as $ \hat d \cdot \hat u = \cos \delta \cos \delta_d 
\cos (\alpha-\alpha_d) + \sin \delta \sin \delta_d$, with  $\delta_d$ the 
dipole declination and $\alpha_d$ its right ascension, and performing the 
integration in $\alpha$ in the previous equations, it can be seen that
\begin{equation}
\label{eqn:amplitudes}
r=\left| \frac{Ad_\perp}{1+Bd_z} \right|
\end{equation}
where 
$$A =\frac{\int d\delta\,\omega(\delta) \cos^2 \delta}
{\int d\delta\,\omega(\delta) \cos \delta}, \hspace{1cm}  
B =\frac{\int d\delta\,\omega(\delta) \cos \delta \sin \delta}
{\int d\delta\,\omega(\delta) \cos \delta}$$ 
and $d_z=d\sin{\delta_d}$ denotes the component of the dipole along the 
Earth rotation axis while $d_\perp=d\cos{\delta_d}$ is the component in the 
equatorial plane \cite{julien}. The coefficients $A$ and $B$ can be
estimated from the data as the mean values of the cosine 
and the sine of the event declinations. In our case, 
$A=\left<\cos{\delta}\right>\simeq 0.78$ and $B=\left<\sin{\delta}\right>\simeq -0.45$. 
For a dipole amplitude $d$, the measured amplitude of the first harmonic
 in right ascension $r$ thus depends on the 
region of the sky observed, which is essentially a function of the latitude 
of the observatory $\ell_{site}$, and the range of zenith angles considered. In the case of a 
small $B d_z$ factor, the dipole  component in the equatorial plane $d_\perp$ 
is obtained as $d_\perp\simeq r/\left<\cos{\delta}\right>$.
The phase $\varphi$ corresponds to the right ascension of the dipole direction
$\alpha_d$.

Turning now to the East-West method, the measured flux from the East sector 
for a local sidereal time $\alpha^0$ can be similarly expressed as
\begin{equation}
I_E (\alpha^0)=\int_{-\pi/2}^{\pi/2} d\phi \int_0^{\theta_{max}} d\theta 
\sin \theta\ \epsilon (\theta) \ J(\theta,\phi,\alpha^0),
\end{equation}
and analogously for the measured flux coming from the west sector
changing the azimuthal integration to the interval 
$[\pi/2,3\pi/2]$. Expressing $\hat d \cdot \hat u$ in local coordinates ($\theta$, $\phi$ 
and $\alpha^0$), and performing the integration over $\phi$ we obtain for the leading order
\begin{equation}
\frac{I_E - I_W}{\langle I_E + I_W \rangle} (\alpha^0) =-\frac{2 d_\perp C 
}{\pi (1+d_z D \sin \ell_{site})} \sin (\alpha^0 -\alpha_d),
\end{equation}
where 
$$C =\frac{\int d\theta\,\epsilon(\theta) \sin^2 \theta
}{\int d\theta\,\epsilon(\theta) \sin \theta},\hspace{1cm} 
D =\frac{\int d\theta\,\epsilon(\theta) \sin \theta \cos \theta}{
\int d\theta\,\epsilon(\theta) \sin \theta}.$$
In this calculation any dependence of the exposure on the local sidereal time $\alpha^0$ gives 
at first order the same contribution to the East and West sectors flux, and thus gives a negligible 
contribution to the flux difference\footnote{The exposure dependence on azimuth present in Auger 
due to trigger effects at low energies has a $2\pi/6$ frequency that makes it cancel out in the 
computation of the first harmonic, and thus does not affect the above result.}. The next leading 
order, proportional to the equatorial dipole component times the sidereal modulation of the exposure, 
is negligible. The coefficients $C$ and $D$ can be estimated from the observed zenith angles of the 
events. In our case, $C =\left<\sin\theta\right>\simeq 0.58$ and 
$D=\left<\cos\theta\right>\simeq 0.78$. The total detected flux averaged over the local sidereal 
time can be estimated as $\langle I_E +I_W\rangle = N/2\pi$. In case $Dd_z\ll 1$, we get finally~:
\begin{equation}
(I_E - I_W) (\alpha^0) = - \frac{N}{2\pi}\frac{2 d_\perp \langle \sin \theta
\rangle}{\pi} \sin (\alpha^0 - \alpha_d).
\end{equation}

\section*{Acknowledgements}

The successful installation and commissioning of the Pierre Auger Observatory
would not have been possible without the strong commitment and effort
from the technical and administrative staff in Malarg\"ue.

We are very grateful to the following agencies and organizations for financial support: 
Comisi\'on Nacional de Energ\'{\i}a At\'omica, 
Fundaci\'on Antorchas,
Gobierno De La Provincia de Mendoza, 
Municipalidad de Malarg\"ue,
NDM Holdings and Valle Las Le\~nas, in gratitude for their continuing
cooperation over land access, Argentina; 
the Australian Research Council;
Conselho Nacional de Desenvolvimento Cient\'{\i}fico e Tecnol\'ogico (CNPq),
Financiadora de Estudos e Projetos (FINEP),
Funda\c{c}\~ao de Amparo \`a Pesquisa do Estado de Rio de Janeiro (FAPERJ),
Funda\c{c}\~ao de Amparo \`a Pesquisa do Estado de S\~ao Paulo (FAPESP),
Minist\'erio de Ci\^{e}ncia e Tecnologia (MCT), Brazil;
AVCR, AV0Z10100502 and AV0Z10100522,
GAAV KJB300100801 and KJB100100904,
MSMT-CR LA08016, LC527, 1M06002, and MSM0021620859, Czech Republic;
Centre de Calcul IN2P3/CNRS, 
Centre National de la Recherche Scientifique (CNRS),
Conseil R\'egional Ile-de-France,
D\'epartement  Physique Nucl\'eaire et Corpusculaire (PNC-IN2P3/CNRS),
D\'epartement Sciences de l'Univers (SDU-INSU/CNRS), France;
Bundesministerium f\"ur Bildung und Fors-
chung (BMBF),
Deutsche Forschungsgemeinschaft (DFG),
Finanzministerium Baden -
W\"urttemberg,
Helmholtz-Gemeinschaft Deutscher Forschungszentren (HGF),
Ministerium f\"ur Wissenschaft und Forschung, Nordrhein-Westfalen,
Ministerium f\"ur Wissenschaft, Fors-
chung und Kunst, Baden-W\"urttemberg, Germany; 
Istituto Nazionale di Fisica Nucleare (INFN),
Istituto Nazionale di Astrofisica (INAF),
Ministero dell'Istruzione, dell'Universit\`a e della Ricerca (MIUR), 
Gran Sasso Center for Astroparticle Physics (CFA), Italy;
Consejo Nacional de Ciencia y Tecnolog\'{\i}a (CONACYT), Mexico;
Ministerie van Onderwijs, Cultuur en Wetenschap,
Nederlandse Organisatie voor Wetenschappelijk Onderzoek (NWO),
Stichting voor Fundamenteel Onderzoek der Materie (FOM), Netherlands;
Ministry of Science and Higher Education,
Grant Nos. 1 P03 D 014 30 and N N202 207238, Poland;
Funda\c{c}\~ao para a Ci\^{e}ncia e a Tecnologia, Portugal;
Ministry for Higher Education, Science, and Technology,
Slovenian Research Agency, Slovenia;
Comunidad de Madrid, 
Consejer\'{\i}a de Educaci\'on de la Comunidad de Castilla La Mancha, 
FEDER funds, 
Ministerio de Ciencia e Innovaci\'on and Consolider-Ingenio 2010 (CPAN),
Generalitat Valenciana, 
Junta de Andaluc\'{\i}a, 
Xunta de Galicia, Spain;
Science and Technology Facilities Council, United Kingdom;
Department of Energy, Contract Nos. DE-AC02-07CH11359, DE-FR02-04ER41300,
National Science Foundation, Grant No. 0450696,
The Grainger Foundation USA; 
ALFA-EC / HELEN,
European Union 6th Framework Program,
Grant No. MEIF-CT-2005-025057, 
European Union 7th Framework Program, Grant No. PIEF-GA-2008-220240,
and UNESCO.

\end{document}